\documentclass[preprint,12pt]{elsarticleLJW}
\usepackage{amssymb}
\usepackage{amsthm}
\usepackage{amsmath}
\usepackage{graphicx}
\usepackage{float}
\usepackage{mathrsfs}
\usepackage{multirow}
\usepackage{epstopdf}

\usepackage[left=1in,right=1in,top=1.1in,bottom=1in]{geometry}

\begin{document}

\begin{frontmatter}

\title{European Power Option Pricing with Extended Vasic\v{e}k Interest Rate and Exponential Ornstein-Uhlenbeck Asset Process under Different Market Assumptions}

\author{Jingwei Liu \corref{cor*}}
\ead{liujingwei03@tsinghua.org.cn}
\cortext[cor*]{Corresponding author.}
\address{School of Mathematical Sciences, Beihang University, Beijing, 102206,P.R China}

\begin{abstract}
We propose a general framework of European power option pricing under two different market assumptions about extended Vasic\v{e}k interest rate process and exponential Ornstein-Uhlenbeck asset process with continuous dividend as underlying, in which the Brownian motions involved in Vasic\v{e}k interest rate and exponential Ornstein-Uhlenbeck process are time--dependent correlated in equivalent martingale measure probability space or real-world probability space respectively.
We first develop European power option pricing in two types of payoffs with martingale method under the market assumption that Vasic\v{e}k interest rate and exponential Ornstein-Uhlenbeck process are correlated in equivalent martingale measure probability space. Then, we solve the European power option pricing under the market assumption that Vasic\v{e}k interest rate and exponential Ornstein-Uhlenbeck process are correlated in real--world probability by constructing a Girsannov transform to map real--world probability to risk--neutral equivalent martingale measure. Finally, the European power option pricing formulae are derived with numeraire change and T--forward measure under the above two market assumptions in  a uniform theoretical framework and close formulae expression.

\end{abstract}

\end{frontmatter}

\section{Introduction}

In recent two decades, as extension of Black-Scholes formula of European option (Black,et al,1973), exponential Ornstein-Uhlenbeck  process and Vasic\v{e}k process are involved in modelling the asset and interest rate respectively (Lin,et al., 2000; Wang, et al., 2002; Liu, et al., 2005; Li, et al., 2008; Liu, 2009;  Zhou, et al., 2011;  Wei, et al., 2013; Cao, et al., 2017; Wang, et al., 2019; Vasicek,1977; Hull,et al. 1993; Brigo, et al.,2007; Liu,2009; Orlando, et al.2020. ), and  power option pricing is also attracted more attention (Liu, et al., 2005; Liu, 2009; Cao, et al., 2017; Kim, et al., 2022) .

Lin, et al. (2000) describe stock price with exponential Ornstein-Uhlenbeck process substituting for traditional exponential Brownian motion to price European option.
Wang, et al. (2002) apply  exponential Brownian motion stock process and  Vasic\v{e}k stochastic interest rate to price European call option and reset option.
Liu, et al.(2004) discuss European option  and foreign currency option under Hull--White interest rate and stock price driven by Ornstein—Uhlenback process with continuous dividend.
Li,et al. (2008) derive a general pricing formula for compound call option  when the interest rate is stochastic by the change of measure, and an analytic pricing formula for compound call option  is also given in an extended Vasicek's interest rate framework.
Liu (2009) develop the martingale method of pricing European power option where stochastic interest rate and risk asset’S price are under Vasicek model and exponential Ornstein—Uhlenbeck process model respectively.
Zhou,et al. (2011) derive European option with the assumption that the underlying asset price and interest rate follow diffusion process using change of numeraires. And obtain a case of European option pricing  with assumption that the underlying asset price follows a Geometry Brownian Motion and the interest rate follows a HJM model.
Wei,et al.(2013) derive a general pricing formula for digital power-option by measurement transformation in an extended Vasicek interest rate framework.
Cao,et al.(2017) discuss a class of power European option by using the changes of numeraire method where the interest rate and stock price  are described by Hull-White interest model and  exponential Ornstein-Uhlenbeck process with continuous dividend respectively.
Wang,et al.,(2019) derive the compaund options with changing rules of interest rate and  stock price which are described by  Hull-White model and exponential Ornstein-Uhlenbeck process respectively.

In this paper, we first extend the constant correlation coefficient of standard Brownian motions in Vasicek model and Ornstein—Uhlenbeck process with dividend to time-dependent, and derive European power option pricing with the market assumption that Vasicek model and exponential Ornstein—Uhlenbeck process are under  equivalent martingale measure (liu,2009).
Then, we derive the above European power option pricing with the market assumption that Vasicek model and exponential Ornstein—Uhlenbeck process are under real-world probability space. We propose a novel Girsanov transform to change the real-world probability space to equivalent martingale measure ahead of European power option pricing.
Last, we derive the above two market assumptions European power option pricing with change of numeraire method (Geman,et al.,1995; Shreve, 2004). Especially , we give the close formulae of European power option pricing in the equivalent martingale measure with numeraire change case is different from Cao, et al. (2017) in expression.

The remainder of this paper is organized as follows. In Section 2, we derive the European power option formula while
 Vasicek interest rate is in the equivalent martingale measurement probability space derived from exponential Ornstein-Uhlenbeck  process.
 In Section 3, we derive the European power option formula while Vasicek interest rate and exponential Ornstein-Uhlenbeck  process are in the real--world probability space. In Section 4, we derive the European power option formula with change of numeraire while
 Vasicek interest rate is in the equivalent martingale measument probability space determined by exponential Ornstein-Uhlenbeck  process. In Section 5, we derive the European power option formula with change numeraire  while Vasicek interest rate  and exponential Ornstein-Uhlenbeck process are in the real--world probability space. Finally, in Section 6, we give the concluding remarks.

\section{Vasic\v{e}k interest rate process is under equivalent martingale measurement}
 \label{}
Suppose in a complete continuous frickless financial market, there are one zero-bound and risk asset, for example stock. The stock price process $S_{t}$, $t\geq 0$ satisfies the following exponential Ornstein-Uhlenbeck  process

\begin{equation}
\begin{array}{ll}
dS(t)=(\mu(t)-q(t)- c \ln S(t)) S(t) dt +\sigma_{S}(t) S(t) dB(t)\\
\end{array}
\end{equation}
where,$\mu(t)$ is mean--reversion,$q(t)$ is continuous dividend.  $\mu(t),q(t)$ are time $t$ dependent functions, $c$ is a constant of real number. $B(t)$ is one--dimensional standard Brownian motion defined on filtering probability space $(\Omega,\mathcal{F},\mathcal{F}_{t},P)$, where $\mathcal{F}_{t}=\sigma(B(s),0\leq s\leq t)$ . Exponential Ornstein-Uhlenbeck process takes advantage of mean--reverting and overcoming the growing trend of geometric Brownian motion, a heuristic simulation is shown in Figure 1 with Runge--Kutta method of stochastic differential equations (Kloeden, et al.,1992).
\begin{figure}[!htbp]
\begin{center}
\begin{minipage}{8cm}
\includegraphics[width=\textwidth]{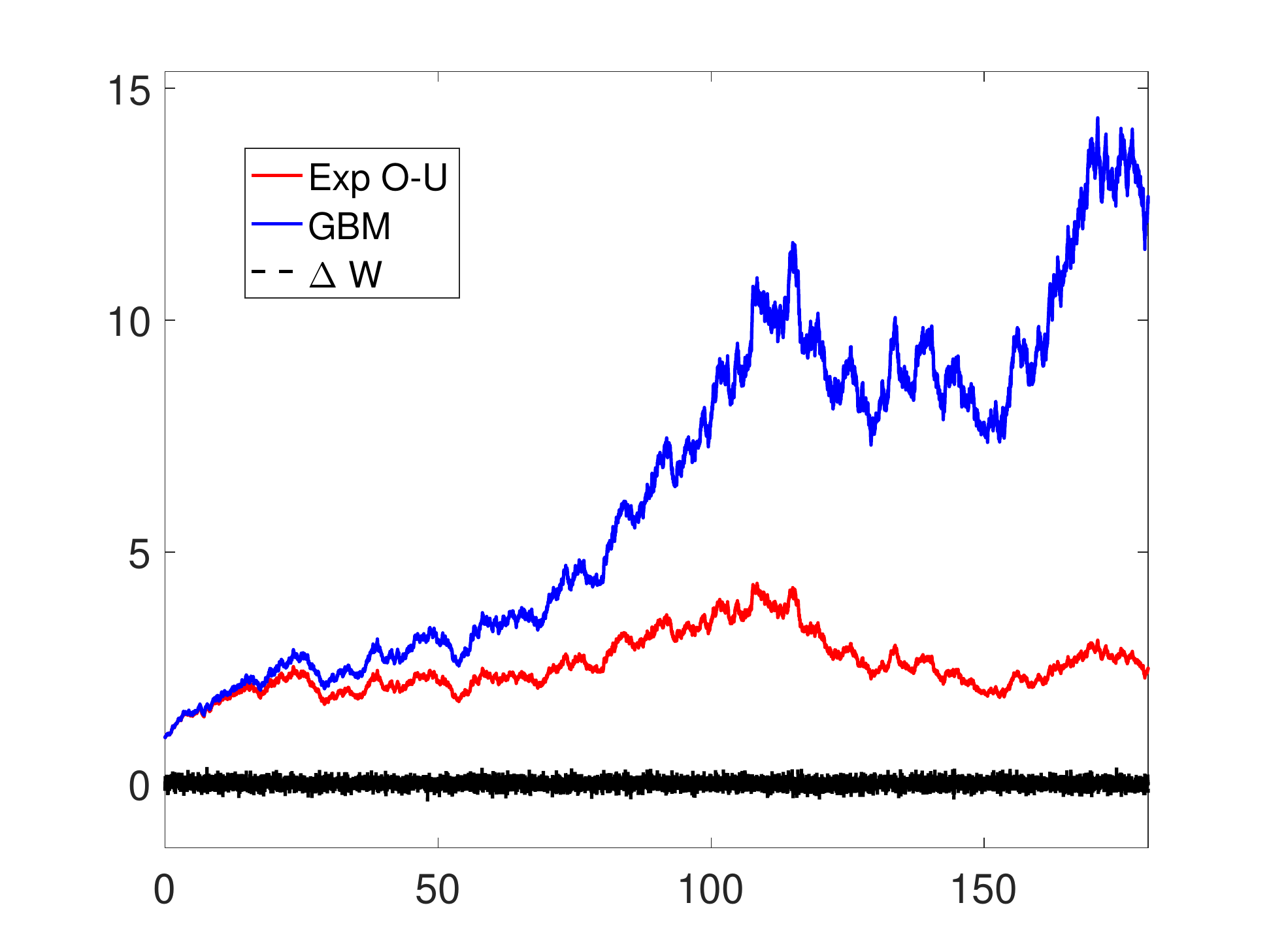}
\caption{Simulation of  Geometric Brownian motion (GBM) and Exponential Ornstein-Uhlenbeck process (Exp O--U) . GBM is with $\mu=0.005$,$q=0$,$c=0$,$\sigma=0.006$, $S(0)=1$. Exp O--U is with $\mu=0.005$,$q=0$,$c=0.01$,$\sigma=0.006$, $S(0)=1$. And $\Delta B=\Delta W \sim N(0,0.01)$.}
\end{minipage}
\end{center}
\end{figure}
Suppose the maturity time is $T$, strike time is $K$, the payoff of  European  call power option takes two types
\begin{equation}
      \xi=(S_{T}^{n}-K^{n})^{+}=\max(S_{T}^{n}-K^{n},0)
\end{equation}
or,
 \begin{equation}
      \xi=(S_{T}^{n}-K)^{+}=\max(S_{T}^{n}-K,0)
      \end{equation}\\
The corresponding payoff functions of  European  put power option are
\begin{equation}
      \eta=(K^{n}-S_{T}^{n})^{+}=\max(K^{n}-S_{T}^{n},0)
      \end{equation}
or,
 \begin{equation}
      \eta=(K-S_{T}^{n})^{+}=\max(K-S_{T}^{n},0)
      \end{equation}\\
where,$n>0,  n\in \mathbf{R}$.  when $n=1$, payoff functions are the standard European call and put option
      \begin{equation}
      \xi=(S_{T}-K)^{+}=\max(S_{T}-K,0) ,\quad  \eta=(K-S_{T})^{+} .
      \nonumber
      \end{equation}

 Suppose asset price process follows exponential Ornstein-Uhlenbeck process, the equivalent martingale measurement is constructed according to Girsanov Theorem and Ito formula as follows.\\

 {\bf Lemma 1} (Lin,et al,2000;Liu,et al,2004; Liu,2009) For  exponential Ornstein-Uhlenbeck stock process (1), let
$ \displaystyle \theta(t) =\displaystyle \frac{\mu(t)-c \ln S(t)-r(t)}{\sigma_{S}}$,
 if
 $\displaystyle \mathbb{E}^{P} \{\exp(\frac{1}{2} \int_{0}^{t} \theta^2(s) ds)\}<\infty$ ,
 there exists a unique equivalent martingale measurement $Q$ satisfying
 \begin{equation}
 \displaystyle \frac{dQ}{dP}=\displaystyle \exp\{-\frac{1}{2} \int_{0}^{T} \theta^2(t) dt - \int_{0}^{T} \theta(t) dB(t)\} ,
 \nonumber
 \end{equation}
such that $\displaystyle S_t^{*}=\displaystyle S_t \exp \{-\int_{0}^{t} (r(t)-q(t)) dt\}$ is $Q-$ martingale. Let
$\displaystyle B^{Q}(t)=\displaystyle B(t)+\int_{0}^{t} \theta(s) ds $, then $B^{Q} (t)$ is Brownian motion under measurement $Q$ , and
 \begin{equation}
 \displaystyle S_{T}=\displaystyle S_{t} \exp\{ \int_{t}^{T} (r(s)-q(s)-\frac{1}{2} \sigma^2_{S}(s)) ds + \int_{t}^{T} \sigma_{S}(s) dB^{Q}(s)\},
 \end{equation}

{\bf Market Assumption 1}: Vasicek interest rate $r(t)$ is under $Q$ measurement,
 \begin{equation}
 \begin{array}{l}
\displaystyle d r_{t}= \displaystyle (\beta_{t}-\alpha_{t}r_{t})dt+ \sigma_{r}(t) dZ^Q(t)      \\
 \end{array}
 \end{equation}
 where $\alpha_{t},\beta_{t},\gamma_{t}$ and $\sigma_{r}(t)$ are time $t$ dependent functions,$Z^Q(t)$ and $B^Q(t)$ are one--dimensional standard Brownian motions with correlation coefficient $\rho_{t}$ defined on  probability space \\
 $(\Omega,\mathcal{F},\{\mathcal{F}_{t}\}_{t\geq 0},Q)$, where $\mathcal{F}_{t}=\sigma(B^Q(s),Z^Q(s),0\leq s\leq t)$. \\

As in Liu,et al.(2004) and Li,et al.(2008), we denote
\begin{equation}
\begin{array}{ll}
    \displaystyle   \lambda(t)=\int_{0}^{t} \alpha_{s}ds, \quad H(u,v)=e^{\{\int_{u}^{v} -\alpha_{s}ds \}}=e^{\lambda(u)-\lambda(v)}, \quad  m(u,v)=\int_{u}^{v} H(u,s) ds, \\
    \displaystyle   G(t,T,r_{t})=r_{t}m(t,T) + \int_{t}^{T} \beta_{s} m(s,T) ds ,\\
    \displaystyle X_{Q}=\int_{t}^{T} \sigma_{r}(s)m(s,T)dZ^Q(s),\quad  Y_{Q}=\int_{t}^{T} \sigma_{S}(s)dB^Q(s).\\
    \sigma_{X_{Q}}^{2}=\int_{t}^{T} \sigma_{r}^{2} (u) m^{2}(u,T)du,  \quad \sigma_{Y_{Q}}^{2}=\int_{t}^{T} \sigma_{S}^{2}(s)ds.
\end{array}
\end{equation}

Applying Ito formula to Vasicek interest rate $r(t)$ in Market Assumption 1, we obtain Lemma 2.

{\bf Lemma 2}. (Liu,et al,2004;Li,et al.2008; Liu,2009) For Vasciek interest rate $r(t)$, $0\leq t\leq T$,
 \begin{equation}
 \begin{array}{ll}
\displaystyle \int_{t}^{T} r_{s} ds=\displaystyle  r_{t}m(t,T) + \int_{t}^{T} \beta_{s} m(s,T) ds+  \int_{t}^{T} \sigma_{r} (s) m(s,T) dZ^Q(s)  \\
 \end{array}
 \end{equation}

The following Lemma 3 is widely used in conditional mathematical expectation of option price calculation.

{\bf Lemma 3}.(Chen,2002) suppose $W_{1}\sim N(0,1)$, $W_{2}\sim N(0,1)$, $Cov{(W_{1},W_{2})}=\rho$, then for any real number $a,b,c,d,k$,

\begin{equation}
\begin{array}{l}
\mathbb{E}[e^{cW_{1}+dW_{2}}I_{\{aW_{1}+bW_{2}\geq k\}}]=e^{\frac{1}{2}(c^2+d^2+2\rho cd)}N(\frac{\displaystyle ac+bd+\rho(ad+bc)-k}{\displaystyle \sqrt{a^2+b^2+2\rho ab}})
 \end{array}
\end{equation}

Then, we have the following two theorems under Market Assumption 1.\\

{\bf Theorem 1. } Suppose $Z^{Q}(t)$ and $B^{Q}(t)$ are two standard one-dimension Brown motions on probability space with correlation coefficient $\rho_t$ $( -1\leq \rho_t \leq 1)$ of exponential Ornstein-Uhlenbeck stock process (1) and Vasicek interest rate (7), with strike price $K$, and maturity time $T$, European power call option price $V_{c}(t,r_{t},S_{t},T,K)$ at time $t (0\leq t\leq T)$ with payoff function (2) is

\begin{equation}
\begin{array}{ll}
V_{c}(t,r_{t},S_{t},T,K)=S_{t}^{n} e^{ (n-1)G(t,T,r_{t})-n\int_{t}^{T} q(s)ds +\frac{(n-1)^2}{2}\sigma_{X_{Q}}^{2}+\frac{n(n-1)}{2} \sigma_{Y_{Q}}^{2}+\rho_t^{'}n(n-1)\sigma_{X_{Q}}\sigma_{Y_{Q}}}N(d_{1})\\
                  \hspace*{2.0cm}  -K^{n} e^{ -G(t,T,r_{t})+\frac{\sigma_{X_{Q}}^{2}}{2}}N(d_{2}).
\end{array}
\end{equation}

European power put option price $V_{p}(t,r_{t},S_{t},T,K)$ at time $t (0\leq t\leq T)$ with payoff function (4) is

     \begin{equation}
     \begin{array}{ll}
      V_{p}(t,r_{t},S_{t},T,K)=K^{n} e^{ -G(t,T,r_{t})+\frac{\sigma_{X_{Q}}^{2}}{2}}N(-d_{2})\\
                         \hspace*{2.0cm} -S_{t}^{n} e^{(n-1)G(t,T,r_{t})-n\int_{t}^{T} q(s)ds +\frac{(n-1)^2}{2}\sigma_{X_{Q}}^{2}+\frac{n(n-1)}{2} \sigma_{Y_{Q}}^{2}+\rho_t^{'}n(n-1)\sigma_{X_{Q}}\sigma_{Y_{Q}}} N(-d_{1}).
     \end{array}
     \end{equation}

European power put--call parity is

    \begin{equation}
    \begin{array}{ll}
    &V_{c}(t,r_{t},S_{t},T,K)+K^{n} e^{ -G(t,T,r_{t})+\frac{\sigma_{X_{Q}}^{2}}{2}}\\
    &=V_{p}(t,r_{t},S_{t},T,K)+S_{t}^{n} e^{ (n-1)G(t,T,r_{t}) -n\int_{t}^{T} q(s)ds+\frac{(n-1)^2}{2}\sigma_{X_{Q}}^{2}+\frac{n(n-1)}{2} \sigma_{Y_{Q}}^{2}+\rho_t^{'}n(n-1)\sigma_{X_{Q}}\sigma_{Y_{Q}}}.
    \end{array}
    \end{equation}
where,

     \begin{equation}
     \begin{array}{ll}
     d_{1}=\frac{\ln\frac{S_{t}}{K}+G(t,T,r_{t})-\int_{t}^{T} q(s)ds+(n-1)\sigma_{X_{Q}}^{2}+(n-\frac{1}{2})\sigma_{Y_{Q}}^{2}+\rho_t^{'}(2n-1)\sigma_{X_{Q}}\sigma_{Y_{Q}}} {\sqrt{\sigma_{X_{Q}}^{2}+\sigma_{Y_{Q}}^{2}+2\rho_t^{'}\sigma_{X_{Q}}\sigma_{Y_{Q}}}},\\
     d_{2}=\frac{ \ln\frac{S_{t}}{K}+G(t,T,r_{t})-\int_{t}^{T} q(s)ds-(\sigma_{X_{Q}}^{2}+\frac{\sigma_{Y_{Q}}^{2}}{2}+\rho_t^{'}\sigma_{X_{Q}}\sigma_{Y_{Q}})} {\sqrt{\sigma_{X_{Q}}^{2}+\sigma_{Y_{Q}}^{2}+2\rho_t^{'}\sigma_{X_{Q}}\sigma_{Y_{Q}}}},\\
     \rho_t^{'}= \frac{ \int_{t}^{T} \rho_u \sigma_{r}(u)\sigma_{u}m(u,T)du}{ \sigma_{X_{Q}}\sigma_{Y_{Q}}}.
     \end{array}
     \end{equation}
where $N(x)$ denotes standard normal distribution.\\

\begin{proof}  According to arbitrage risk-neutral theory , European power call option at time $t$  with  payoff function $\xi=(S_{T}^{n}-K^{n})^{+}$  is
\begin{equation}
\begin{array}{ll}
V_{c}(t,r_{t},S_{t},T,K)&=\mathbb{E^Q}[ (S_{T}^{n}-K^{n})^{+} e^{-\int_{t}^{T} r(s)ds}|\mathcal{F}_{t}]\\
                      &=\mathbb{E^Q}[S_{T}^{n} e^{-\int_{t}^{T} r(s)ds} I_{\{S_{T}^{n}>K^{n}\}} |\mathcal{F}_{t}]-K^{n}\mathbb{E^Q}[e^{-\int_{t}^{T} r(s)ds}I_{\{S_{T}^{n}>K^{n}\}}|\mathcal{F}_{t}]\\
                      &\stackrel{\bigtriangleup}{=} A-B
\end{array}
\nonumber
\end{equation}

Since,
\begin{equation}
\rho_t^{'}=\frac{Cov(\frac{X_Q}{\sigma_{X_Q}},\frac{Y_Q}{\sigma_{Y_Q}})}{ \sqrt{D(\frac{X_Q}{\sigma_{X_Q}})}\sqrt{D(\frac{Y_Q}{\sigma_{Y_Q}}})}
         =\frac{ {Cov(X_Q,Y_Q)}}{ \sigma_{X_Q}\sigma_{Y_Q}}= \frac{ \int_{t}^{T}   \rho_u \sigma_{r}(u)\sigma_{S}(u)m(u,T)du}{ \sigma_{X_Q}\sigma_{Y_Q}},
\nonumber
\end{equation}
combing formulae (6)(8)(9) and Lemma 3, we obtain
\begin{equation}
\begin{array}{ll}
A&= \mathbb{E^Q}[S_{t}^{n} e^{(n-1)\int_{t}^{T} r(s)ds -n\int_{t}^{T} q(s)ds -\int_{t}^{T}\frac{n}{2} \sigma_{S}^{2}(s)ds + \int_{t}^{T} n\sigma_{S}(s)dB^Q(s)}I_{\{\int_{t}^{T} (r(s)-q(s)-\frac{1}{2} \sigma_{S}^{2}(s))ds + \int_{t}^{T} \sigma_{S}(s)dB^Q(s)>\ln\frac{K}{S_{t}}\}} |\mathcal{F}_{t}]\\
 &= \mathbb{E^Q}[S_{t}^{n} e^{(n-1)(G(t,T,r_{t})+X_Q)-n\int_{t}^{T} q(s)ds -\int_{t}^{T}\frac{n}{2} \sigma_{S}^{2}(s)ds + nY_Q } I_{\{G(t,T,r_{t})+X_Q-\int_{t}^{T} q(s)ds-\int_{t}^{T} \frac{1}{2} \sigma_{S}^{2}(s)ds+Y_Q>\ln\frac{K}{S_{t}}\}} |\mathcal{F}_{t}]\\
 &= S_{t}^{n} e^{(n-1)G(t,T,r_{t})-n\int_{t}^{T} q(s)ds -\int_{t}^{T}\frac{n}{2} \sigma_{S}^{2}(s)ds }
 \mathbb{E^Q}[e^{(n-1)X_Q+ nY_Q} I_{\{X_Q+Y_Q>\ln\frac{K}{S_{t}}-G(t,T,r_{t})+\int_{t}^{T} q(s)ds+\int_{t}^{T}\frac{1}{2} \sigma_{S}^{2}(s)ds \}} |\mathcal{F}_{t}] \\
 &= S_{t}^{n} e^{(n-1)G(t,T,r_{t})-n\int_{t}^{T} q(s)ds -\frac{n}{2} \sigma_{Y_Q}^{2}} \mathbb{E}[e^{(n-1)X_Q+ nY_Q} I_{\{X_Q+Y_Q>\ln\frac{K}{S_{t}}-G(t,T,r_{t})+\int_{t}^{T} q(s)ds+\frac{1}{2} \sigma_{Y_Q}^{2}\}} |\mathcal{F}_{t}]\\
 &= S_{t}^{n} e^{(n-1)G(t,T,r_{t})-n\int_{t}^{T} q(s)ds -\frac{n}{2} \sigma_{Y_Q}^{2}} e^{\frac{1}{2}[(n-1)^2 \sigma_{X_Q}^{2}+n^2 \sigma_{Y_Q}^{2}+2\rho_t^{'}n(n-1)\sigma_{X_Q}\sigma_{Y_Q}]}\\
 & \quad  N\Big( \frac{\ln\frac{S_{t}}{K}+G(t,T,r_{t})-\int_{t}^{T} q(s)ds+(n-1)\sigma_{X_Q}^{2}+(n-\frac{1}{2})\sigma_{Y_Q}^{2}+\rho_t^{'}(2n-1)\sigma_{X_Q}\sigma_{Y_Q}} { \sqrt{\sigma_{X_Q}^{2}+\sigma_{Y_Q}^{2}+2\rho_t^{'}\sigma_{X_Q}\sigma_{Y_Q}}}\Big)\\
 &= S_{t}^{n} e^{(n-1)G(t,T,r_{t})-n\int_{t}^{T} q(s)ds +\frac{(n-1)^2}{2}\sigma_{X_Q}^{2}+\frac{n(n-1)}{2} \sigma_{Y_Q}^{2}+\rho_t^{'}n(n-1)\sigma_{X_Q}\sigma_{Y_Q}}\\
 & \quad  N\Big(\frac{\ln\frac{S_{t}}{K}+G(t,T,r_{t})-\int_{t}^{T} q(s)ds +(n-1)\sigma_{X_Q}^{2}+(n-\frac{1}{2})\sigma_{Y_Q}^{2}+\rho_t^{'}(2n-1)\sigma_{X_Q}\sigma_{Y_Q}} { \sqrt{\sigma_{X_Q}^{2}+\sigma_{Y_Q}^{2}+2\rho_t^{'}\sigma_{X_Q}\sigma_{Y_Q}}}\Big)\\
\end{array}
\nonumber
\end{equation}

\begin{equation}
\begin{array}{ll}
B&= K^{n}\mathbb{E^Q}[e^{-G(t,T,r_{t})-X_Q}I_{\{X_Q+Y_Q>\ln\frac{K}{S_{t}}-G(t,T,r_{t})+\int_{t}^{T} q(s)ds+ \frac{1}{2} \sigma_{Y_Q}^{2}\}} |\mathcal{F}_{t}]\\
 &=K^{n} e^{-G(t,T,r_{t})}\mathbb{E}[e^{-X_Q}I_{\{X_Q+Y_Q>\ln\frac{K}{S_{t}}-G(t,T,r_{t})+\int_{t}^{T} q(s)ds+ \frac{1}{2} \sigma_{Y_Q}^{2}\}} |\mathcal{F}_{t}]\\
 &= K^{n} e^{ -G(t,T,r_{t})+ \frac{1}{2}\sigma_{X_Q}^{2} } N\Big( \frac{\ln\frac{S_{t}}{K}+G(t,T,r_{t})-\int_{t}^{T} q(s)ds-(\sigma_{X_Q}^{2}+\frac{\sigma_{Y_Q}^{2}}{2}+\rho_t^{'}\sigma_{X_Q}\sigma_{Y_Q})} {\sqrt{\sigma_{X_Q}^{2}+\sigma_{Y_Q}^{2}+2\rho_t^{'}\sigma_{X_Q}\sigma_{Y_Q}}}\Big)
\end{array}
\nonumber
\end{equation}
Hence formula (11) is proved.
Similarly,  formula (12) can be derived. According to the property of normal distribution $N(-x)=1-N(x)$, formula (13) can be easily verified. Therefore, it completes the proof.
\end{proof}

{\bf Theorem 2. } Suppose $Z^{Q}(t)$ and $B^{Q}(t)$ are two standard one-dimension Brown motions on probability space with correlation coefficient $\rho_t$ $( -1\leq \rho_t \leq 1)$ of exponential Ornstein-Uhlenbeck stock process (1) and Vasicek interest rate (7),  with strike price $K$, and maturity time $T$, the European power call option price  $V_{c}(t,r_{t},S_{t},T,K)$ at time $t$ $(0\leq t\leq T)$ with payoff function (3) is

\begin{equation}
\begin{array}{ll}
V_{c}(t,r_{t},S_{t},T,K)= S_{t}^{n} e^{ (n-1)G(t,T,r_{t})-n\int_{t}^{T} q(s)ds +\frac{(n-1)^2}{2}\sigma_{X_{Q}}^{2}+\frac{n(n-1)}{2} \sigma_{Y_{Q}}^{2}+\rho_t^{'}n(n-1)\sigma_{X_{Q}}\sigma_{Y_{Q}}}N(d_{1})\\
                  \hspace*{2.0cm}  -K e^{ -G(t,T,r_{t})+\frac{\sigma_{X_{Q}}^{2}}{2}}N(d_{2}).
\end{array}
\end{equation}

European power put option $V_{p}(t,r_{t},S_{t},T,K)$ at time $t (0\leq t\leq T)$ with payoff function (5) is

\begin{equation}
\begin{array}{ll}
V_{p}(t,r_{t},S_{t},T,K)= K e^{ -G(t,T,r_{t})+\frac{\sigma_{X_{Q}}^{2}}{2}}N(-d_{2})\\
                         \hspace*{2.0cm} -S_{t}^{n} e^{ (n-1)G(t,T,r_{t})-n\int_{t}^{T} q(s)ds +\frac{(n-1)^2}{2}\sigma_{X_{Q}}^{2}+\frac{n(n-1)}{2} \sigma_{Y_{Q}}^{2}+\rho_t^{'}n(n-1)\sigma_{X_{Q}}\sigma_{Y_{Q}}} N(-d_{1}).
\end{array}
\end{equation}

European power option put--call parity is
\begin{equation}
    \begin{array}{ll}
    &V_{c}(t,r_{t},S_{t},T,K)+K e^{ -G(t,T,r_{t})+\frac{\sigma_{X_{Q}}^{2}}{2}}\\
    &=V_{p}(t,r_{t},S_{t},T,K)+S_{t}^{n} e^{ (n-1)G(t,T,r_{t})-n\int_{t}^{T} q(s)ds +\frac{(n-1)^2}{2}\sigma_{X_{Q}}^{2}+\frac{n(n-1)}{2} \sigma_{Y_{Q}}^{2}+\rho_t^{'}n(n-1)\sigma_{X_{Q}}\sigma_{Y_{Q}}}.
    \end{array}
    \end{equation}

where,
 \begin{equation}
     \begin{array}{ll}
     d_{1}=\frac{\ln\frac{S_{t}}{\sqrt[n]{K}}+G(t,T,r_{t})-\int_{t}^{T} q(s)ds+(n-1)\sigma_{X_{Q}}^{2}+(n-\frac{1}{2})\sigma_{Y_{Q}}^{2}+\rho_t^{'}(2n-1)\sigma_{X_{Q}}\sigma_{Y_{Q}}}
     {\sqrt{\sigma_{X_{Q}}^{2}+\sigma_{Y_{Q}}^{2}+2\rho_t^{'}\sigma_{X_{Q}}\sigma_{Y_{Q}}}},\\
     d_{2}=\frac{\ln\frac{S_{t}}{\sqrt[n]{K}}+G(t,T,r_{t})-\int_{t}^{T} q(s)ds-(\sigma_{X_{Q}}^{2}+\frac{\sigma_{Y_{Q}}^{2}}{2}+\rho_t^{'}\sigma_{X_{Q}}\sigma_{Y_{Q}})}
     {\sqrt{\sigma_{X_{Q}}^{2}+\sigma_{Y_{Q}}^{2}+2\rho_t^{'}\sigma_{X_{Q}}\sigma_{Y_{Q}}}},\\
     \rho_t^{'}= \frac{ \int_{t}^{T} \rho_u \sigma_{r}(u)\sigma_{u}m(u,T)du}{\sigma_{X_{Q}}\sigma_{Y_{Q}}}.
     \end{array}
     \end{equation}

where, $N(x)$ denotes standard normal distribution. \\

\begin{proof} According to arbitrage risk-neutral theory, the price of  European power call option with  payoff function (3) at time $t$ is
\begin{equation}
\begin{array}{ll}
V_{c}(t,r_{t},S_{t},T,K)&=\mathbb{E^Q}[ (S_{T}^{n}-K)^{+} e^{-\int_{t}^{T} r(s)ds}|\mathcal{F}_{t}]\\
                      &=\mathbb{E^Q}[S_{T}^{n} e^{-\int_{t}^{T} r(s)ds}I_{\{S_{T}^{n}>K\}} |\mathcal{F}_{t}]-K \mathbb{E^Q}[e^{-\int_{t}^{T} r(s)ds}I_{\{S_{T}^{n}>K\}}|\mathcal{F}_{t}]\\
                      &\stackrel{\bigtriangleup}{=} A-B
\end{array}
\nonumber
\end{equation}

Since,
\begin{equation}
\rho_t^{'}=\frac{  Cov(\frac{X_Q}{\sigma_{X_Q}},\frac{Y_Q}{\sigma_{Y_Q}})}{\sqrt{D(\frac{X_Q}{\sigma_{X_Q}})}\sqrt{D(\frac{Y_Q}{\sigma_{Y_Q}}})}
         =\frac{{Cov(X_Q,Y_Q)}}{ \sigma_{X_Q}\sigma_{Y_Q}}
         =\frac{\int_{t}^{T} \rho_u\sigma_{r}(u)\sigma_{u}m(u,T)du}{\sigma_{X_Q}\sigma_{Y_Q}},
\nonumber
\end{equation}
combining formulae (6)(8)(9) and Lemma 3, we obtain

\begin{equation}
\begin{array}{ll}
A&=\mathbb{E^Q}[S_{t}^{n} e^{(n-1)\int_{t}^{T} r(s)ds -n\int_{t}^{T} q(s)ds-\int_{t}^{T}\frac{n}{2} \sigma_{S}^{2}(s)ds + \int_{t}^{T} n\sigma_{S}(s)dB_{s}}I_{\{\int_{t}^{T} (r(s)-q(s)-\frac{1}{2} \sigma_{S}^{2}(s))ds + \int_{t}^{T} \sigma_{S}(s)dB_{s}>\ln\frac{\sqrt[n]{K}}{S_{t}}\}} |\mathcal{F}_{t}]\\
 &=\mathbb{E^Q}[S_{t}^{n} e^{(n-1)(G(t,T,r_{t})+X_Q)-n\int_{t}^{T} q(s)ds -\int_{t}^{T}\frac{n}{2} \sigma_{S}^{2}(s)ds + nY_Q } I_{\{G(t,T,r_{t})+X_Q-\int_{t}^{T} q(s)ds-\int_{t}^{T}\frac{1}{2} \sigma_{S}^{2}(s)ds+Y_Q>\ln\frac{\sqrt[n]{K}}{S_{t}}\}} |\mathcal{F}_{t}]\\
 &= S_{t}^{n} e^{(n-1)G(t,T,r_{t})-n\int_{t}^{T} q(s)ds -\int_{t}^{T}\frac{n}{2} \sigma_{S}^{2}(s)ds } \mathbb{E^Q}[e^{(n-1)X_Q+ nY_Q}
 I_{\{X_Q+Y_Q>\ln\frac{\sqrt[n]{K}}{S_{t}}-G(t,T,r_{t})+\int_{t}^{T} q(s)ds+\int_{t}^{T}\frac{1}{2} \sigma_{S}^{2}(s)ds \}} |\mathcal{F}_{t}]\\
 &= S_{t}^{n} e^{(n-1)G(t,T,r_{t})-n\int_{t}^{T} q(s)ds -\frac{n}{2} \sigma_{Y_Q}^{2}} \mathbb{E^Q}[e^{(n-1)X_Q+ nY_Q}I_{\{X_Q+Y_Q>\ln\frac{\sqrt[n]{K}}{S_{t}}-G(t,T,r_{t})+\int_{t}^{T} q(s)ds+\frac{1}{2} \sigma_{Y_Q}^{2}\}} |\mathcal{F}_{t}]\\
 &= S_{t}^{n} e^{(n-1)G(t,T,r_{t}) -n\int_{t}^{T} q(s)ds-\frac{n}{2} \sigma_{Y_Q}^{2}} e^{\frac{1}{2}[(n-1)^2 \sigma_{X_Q}^{2}+n^2 \sigma_{Y_Q}^{2}+2\rho_t^{'}n(n-1)\sigma_{X_Q}\sigma_{Y_Q}]}\\
 & \quad  N\Big(\frac{\ln\frac{S_{t}}{\sqrt[n]{K}}+G(t,T,r_{t})-\int_{t}^{T} q(s)ds+(n-1)\sigma_{X_Q}^{2}+(n-\frac{1}{2})\sigma_{Y_Q}^{2}+\rho_t^{'}(2n-1)\sigma_{X}\sigma_{Y}} { \sqrt{\sigma_{X_Q}^{2}+\sigma_{Y_Q}^{2}+2\rho_t^{'}\sigma_{X_Q}\sigma_{Y_Q}}}\Big)\\
 &= S_{t}^{n} e^{(n-1)G(t,T,r_{t})-n\int_{t}^{T} q(s)ds +\frac{(n-1)^2}{2}\sigma_{X_Q}^{2}+\frac{n(n-1)}{2} \sigma_{Y_Q}^{2}+\rho_t^{'}n(n-1)\sigma_{X_Q}\sigma_{Y_Q}}\\
 & \quad  N\Big(\frac{\ln\frac{S_{t}}{\sqrt[n]{K}}+G(t,T,r_{t})-\int_{t}^{T} q(s)ds +(n-1)\sigma_{X_Q}^{2}+(n-\frac{1}{2})\sigma_{Y_Q}^{2}+\rho_t^{'}(2n-1)\sigma_{X_Q}\sigma_{Y_Q}} { \sqrt{\sigma_{X_Q}^{2}+\sigma_{Y_Q}^{2}+2\rho_t^{'}\sigma_{X_Q}\sigma_{Y_Q}}}\Big)\\
\end{array}
\nonumber
\end{equation}

\begin{equation}
\begin{array}{ll}
B&= K\mathbb{E^Q}[e^{-G(t,T,r_{t})-X_Q}I_{\{X_Q+Y_Q>\ln\frac{\sqrt[n]{K}}{S_{t}}-G(t,T,r_{t})+\int_{t}^{T} q(s)ds+ \frac{1}{2} \sigma_{Y_Q}^{2}\}} |\mathcal{F}_{t}]\\
 &=K e^{-G(t,T,r_{t})}\mathbb{E^Q}[e^{-X_Q}I_{\{X_Q+Y_Q>\ln\frac{\sqrt[n]{K}}{S_{t}}-G(t,T,r_{t})+\int_{t}^{T} q(s)ds+ \frac{1}{2} \sigma_{Y_Q}^{2}\}} |\mathcal{F}_{t}]\\
 &= K e^{-G(t,T,r_{t})+\frac{1}{2}\sigma_{X_Q}^{2} } N\Big(\frac{\ln\frac{S_{t}}{\sqrt[n]{K}}+G(t,T,r_{t})-\int_{t}^{T} q(s)ds-(\sigma_{X_Q}^{2}+\frac{\sigma_{Y_Q}^{2}}{2}+\rho_t^{'}\sigma_{X_Q}\sigma_{Y_Q})} {\sqrt{\sigma_{X_Q}^{2}+\sigma_{Y_Q}^{2}+2\rho_t^{'}\sigma_{X_Q}\sigma_{Y_Q}}}\Big)
\end{array}
\nonumber
\end{equation}

Hence, formulae (15) is proved. And, Formula (16) can be derived similarly. According to the property of normal distribution $N(-x)=1-N(x)$, formula (17) can be easily verified. Therefore, it completes the proof. \\
\end{proof}

\section{Vasicek interest rate process and exponential Ornstein-Uhlenbeck process are in real--world probability space }
In this section, we discuss the European power option pricing with assumption that Vasicek interest rate $r(t)$ and exponential O-UOrnstein-Uhlenbeck process are in the real--world  probability space, we denote it  {\bf Market Assumption 2}:
Suppose in a complete continuous frictionless financial market, there are one zero-bound and risk asset for example stock. The short term interest rate $r_{t}$ and risk price process $S_{t}$  ($t\geq 0$) satisfy the following stochastic differential equation
\begin{equation}
\left\{\begin{array}{l}
\displaystyle dS(t)=(\mu(t)-q(t)- c \ln S(t)) S(t) dt +\sigma_{S}(t) S(t) dB(t)\\
\displaystyle d r_{t}= \displaystyle (\beta_{t}-\alpha_{t}r_{t})dt+ \sigma_{r}(t) dZ(t) \\
\end{array}\right.
\end{equation}
where, $\alpha_{t},\beta_{t},\mu(t),q(t)$ and $\sigma_{t}$ are time $t$ dependent functions, $c$ is a constant of real number, $Z(t)$ and $B(t)$ are one--dimensional standard Brownian motions  defined on  probability space $(\Omega,\mathcal{F},\{\mathcal{F}_{t}\}_{t\geq 0},P)$ with correlation coefficient $\rho_t$ $( -1\leq \rho_t \leq 1)$, where $\mathcal{F}_{t}=\sigma(B(s),Z(s),0\leq s\leq t)$. \\

{\bf Lemma 4.} (Shreve, 2004) Suppose $Z(t)$ and $B(t)$ are one--dimensional standard Brownian motions with correlation coefficients $\rho_t$  $( -1\leq \rho_t \leq 1)$ defined on  probability space $(\Omega,\mathcal{F},P)$. There exists a  one--dimensional standard Brownian motion $B^{\perp}(t)$ , $B(t)$ and $B^{\perp}(t)$ are independent, and
       \begin{equation}
       \displaystyle dZ(t)=\rho_t d B(t) +\sqrt{1-\rho_t^2}  dB^{\perp}(t)\\
       \end{equation}

The market processes change to the following stochastic differential equation
\begin{equation}
\left\{\begin{array}{l}
      \displaystyle dS(t)=(\mu(t)-q(t)- c \ln S(t)) S(t) dt +\sigma_{S}(t) S(t) dB(t)\\
      \displaystyle d r_{t}= \displaystyle (\beta_{t}-\alpha_{t}r_{t})dt+ \rho_t \sigma_{r}(t)d B(t) + \sqrt{1-\rho_t^2} \sigma_{r}(t)  dB^{\perp}(t)   \\
\end{array}\right.
\end{equation}

According to 2--dimensional Girsanov Theorem, we construct a Girsanov transform to map real--world space to risk--neutral measurement in Lemma 5.\\

{\bf Lemma 5.}  For $-1\leq \rho_t\leq 1 $ in Market Assumption 2,\\
(1) When  $ |\rho_t|\neq 1 $, let
\begin{equation}
\left\{\begin{array}{l}
\displaystyle \theta_1(t) =\displaystyle \frac{\mu(t)-c \ln S(t)-r(t)}{\sigma_{S}(t)} \\
\displaystyle \theta_2(t)=\displaystyle - \frac{\mu(t)-c \ln S(t)-r(t)}{\sqrt{1-\rho_t^2}\sigma_{S}(t) \sigma_{r}(t) } \rho_t\sigma_{r}(t)  \\
\end{array}\right.
\end{equation}
Denote
 \begin{equation}
 \left\{\begin{array}{l}
  \Theta(t)=(\theta_1(t), \theta_2(t)), \\
  B^Q(t)=B(t)+\int_{0}^{t} \theta_1(u) du ,  \\
  B^{\perp,Q}(t)=B^{\perp}(t)+\int_{0}^{t} \theta_2(u) du
 \end{array}\right.
 \end{equation}
 Let
 \begin{equation}
 \begin{array}{l}
 \mathbb{Z}(t)=\exp\{-\int_{0}^{t} \theta_1(u) dB^Q(u) -\int_{0}^{t}\theta_2(u) d B^{\perp,Q}(u)-\frac{1}{2} \int_{0}^{t}(\theta^2_1(u)+\theta^2_2(u) )du \}    \\
 Q(A)=\int_{A} \mathbb{Z}(T) dP ,   \forall A \in \mathcal{F}
 \end{array}
 \nonumber
 \end{equation}
If $ \mathbb{E}^{P} [\exp(\frac{1}{2} \int_{0}^{t} \|\Theta\|^2(s) ds)]<\infty$,  $(B^Q(t),B^{\perp,Q}(t))$ are 2--dimensional standard Brownian motions under measure $Q$. \\

(2) When  $ |\rho_t|= 1 $, let $\displaystyle \theta_1(t) =\displaystyle \frac{\mu(t)-c \ln S(t)-r(t)}{\sigma_{S}(t)}$ ,
if $ E^{P} [\exp(\frac{1}{2} \int_{0}^{t} \theta_1^2(s) ds)]<\infty$ , there exists a unique equivalent martingale measurement $Q_{S}$, satisfying
 \begin{equation}
 \displaystyle \frac{dQ_S}{dP}=\displaystyle \exp(-\frac{1}{2} \int_{0}^{T} \theta_1^2(t) dt - \int_{0}^{T} \theta_1(t) dB(t)) ,
 \nonumber
 \end{equation}
such that $\displaystyle S_t^{*}=\displaystyle S_t \exp \{-\int_{0}^{t} (r(t)-q(t)) dt\}$ is $Q-$ martingale. Let $\displaystyle B^{Q} (t)=\displaystyle B(t)+\int_{0}^{t} \theta(s) ds $, then, $B^{Q} (t)$ is a Brownian motion under measurement $Q_S$ , and
\begin{equation}
\displaystyle S_{T}=\displaystyle S_{t} \exp( \int_{t}^{T} (r(s)-q(s)-\frac{1}{2} \sigma^2_{S}(s)) ds + \int_{t}^{T} \sigma_{S}(s) dB^{Q}(s)).
\nonumber
\end{equation}  \\
According to Lemma 5, under the equivalent martingale measurement $Q$ , we obtain
\begin{equation}
\left\{\begin{array}{l}
      \displaystyle dS(t)=(r(t)-q(t)) S(t) dt +\sigma_{S}(t) S(t) dB^Q(t)\\
      \displaystyle d r_{t}= \displaystyle (\beta_{t}-\alpha_{t}r_{t})dt+\rho_t \sigma_{r}(t)  d B^Q(t) + \sqrt{1-\rho_t^2} \sigma_{r}(t)  dB^{\perp,Q}(t)   \\
\end{array}\right.
\end{equation}
where $ |\rho_t|\leq 1 $.   \\

Since $B^Q(t)$ and $B^{\perp,Q}(t)$ are independent, $dB^Q(t) d B^{\perp,Q}(t)=0$. We denote
\begin{equation}
\begin{array}{ll}
       \lambda(t)=\int_{0}^{t} \alpha_{u}d u, H(u,v)=e^{\{\int_{u}^{v} -\alpha_{u}d u\}}=e^{\lambda(u)-\lambda(v)},\quad  m(u,v)=\int_{u}^{v} H(u,s) ds, \\
       G(t,T,r_{t})=r_{t}m(t,T) + \int_{t}^{T} \beta_{u}m(u,T)du,\\
       \tilde{X}_Q=\int_{t}^{T} \rho_u \sigma_{r}(u) m(u,T)dB^Q(u)+\int_{t}^{T}  \sqrt{1-\rho_u^2} \sigma_{r}(u)m(u,T)dB^{\perp,Q}(u)\\
       \tilde{Y}_Q=\int_{t}^{T} \sigma_{S}(u)dB^Q(u),\\
       \sigma_{\tilde{X}_Q}^{2}=\int_{t}^{T} \sigma_{r}^{2}(u)m^{2}(u,T)du,  \quad \sigma_{\tilde{Y}_Q}^{2}=\int_{t}^{T} \sigma_{S}^{2}(u)du    \\
\end{array}
\end{equation}

Applying Ito formula to Vasicek interest rate process $r(t)$ in formula (24), we obtain Lemma 6. \\

{\bf Lemma 6.} For Vasicek interest rate process $r(t)$ in formula (24),
\begin{equation}
\begin{array}{ll}
\int_{t}^{T}r(s)ds&=G(t,T,r_{t})+  \int_{t}^{T} \rho_t \sigma_{r}(u)m(u,T)dB^Q(u)\\
                  &+ \int_{t}^{T}\sqrt{1-\rho_t^2}  \sigma_{r}(u)m(u,T)dB^{\perp,Q}(u),\\
\end{array}
\end{equation}

We obtain European power option formulae as follows.\\

{\bf Theorem 3. } Suppose $Z(t)$ and $B(t)$  are two standard one-dimension Brown motions on probability space with correlation coefficient $\rho_t$ $( -1\leq\rho_t \leq 1)$,  interest rate and stock process satisfy stochastic differential equations (19)(21)(24), with strike price $K$, and maturity time $T$, European power call option price $V_{c}(t,r_{t},S_{t},T,K)$ at time $t (0\leq t\leq T)$ with payoff function (2) is

\begin{equation}
\begin{array}{ll}
V_{c}(t,r_{t},S_{t},T,K)=S_{t}^{n} e^{ (n-1)G(t,T,r_{t})-n\int_{t}^{T} q(s)ds +\frac{(n-1)^2}{2}\sigma_{\tilde{X}_{Q}}^{2}+\frac{n(n-1)}{2} \sigma_{\tilde{Y}_{Q}}^{2}+\rho_t^{'}n(n-1)\sigma_{\tilde{X}_{Q}}\sigma_{\tilde{Y}_{Q}}}N(d_{1})\\
                  \hspace*{2.0cm}  -K^{n} e^{ -G(t,T,r_{t})+\frac{\sigma_{\tilde{X}_{Q}}^{2}}{2}}N(d_{2})
\end{array}
\end{equation}

European power put option price $V_{p}(t,r_{t},S_{t},T,K)$ at time $t (0\leq t\leq T)$ with payoff function (4) is

     \begin{equation}
     \begin{array}{ll}
      V_{p}(t,r_{t},S_{t},T,K)=K^{n} e^{ -G(t,T,r_{t})+\frac{\sigma_{\tilde{X}_{Q}}^{2}}{2}}N(-d_{2})\\
                         \hspace*{2.0cm} -S_{t}^{n} e^{(n-1)G(t,T,r_{t})-n\int_{t}^{T} q(s)ds +\frac{(n-1)^2}{2}\sigma_{\tilde{X}_{Q}}^{2}+\frac{n(n-1)}{2} \sigma_{\tilde{Y}_{Q}}^{2}+\rho_t^{'}n(n-1)\sigma_{\tilde{X}_{Q}}\sigma_{\tilde{Y}_{Q}}} N(-d_{1})
     \end{array}
     \end{equation}

European power put--call parity is

    \begin{equation}
    \begin{array}{ll}
    &V_{c}(t,r_{t},S_{t},T,K)+K^{n} e^{ -G(t,T,r_{t})+\frac{\sigma_{\tilde{X}_{Q}}^{2}}{2}}\\
    &=V_{p}(t,r_{t},S_{t},T,K)+S_{t}^{n} e^{ (n-1)G(t,T,r_{t}) -n\int_{t}^{T} q(s)ds+\frac{(n-1)^2}{2}\sigma_{\tilde{X}_{Q}}^{2}+\frac{n(n-1)}{2} \sigma_{\tilde{Y}_{Q}}^{2}+\rho_t^{'}n(n-1)\sigma_{\tilde{X}_{Q}}\sigma_{\tilde{Y}_{Q}}}
    \end{array}
    \end{equation}
where,

     \begin{equation}
     \begin{array}{ll}
     d_{1}=\frac{\ln\frac{S_{t}}{K}+G(t,T,r_{t})-\int_{t}^{T} q(s)ds+(n-1)\sigma_{\tilde{X}_{Q}}^{2}+(n-\frac{1}{2})\sigma_{\tilde{Y}_{Q}}^{2}+\rho_t^{'}(2n-1)\sigma_{\tilde{X}_{Q}}\sigma_{\tilde{Y}_{Q}}} {\sqrt{\sigma_{\tilde{X}_{Q}}^{2}+\sigma_{\tilde{Y}_{Q}}^{2}+2\rho_t^{'}\sigma_{\tilde{X}_{Q}}\sigma_{\tilde{Y}_{Q}}}},\\
     d_{2}=\frac{ \ln\frac{S_{t}}{K}+G(t,T,r_{t})-\int_{t}^{T} q(s)ds -(\sigma_{\tilde{X}_{Q}}^{2}+\frac{\sigma_{\tilde{Y}_{Q}}^{2}}{2}+\rho_t^{'}\sigma_{\tilde{X}_{Q}}\sigma_{\tilde{Y}_{Q}})} {\sqrt{\sigma_{\tilde{X}_{Q}}^{2}+\sigma_{\tilde{Y}_{Q}}^{2}+2\rho_t^{'}\sigma_{\tilde{X}_{Q}}\sigma_{\tilde{Y}_{Q}}}},\\
     \rho_t^{'}= \frac{ \int_{t}^{T} \rho_u \sigma_{r}(u)\sigma_{u}m(u,T)du}{ \sigma_{\tilde{X}_{Q}}\sigma_{\tilde{Y}_{Q}}}
     \end{array}
     \end{equation}

where $N(x)$ denotes standard normal distribution.\\

\begin{proof}   According to arbitrage risk-neutral theory , European power call option at time $t$  with  payoff function $\xi=(S_{T}^{n}-K^{n})^{+}$  is
\begin{equation}
\begin{array}{ll}
V_{c}(t,r_{t},S_{t},T,K)&=\mathbb{E^Q}[ (S_{T}^{n}-K^{n})^{+} e^{-\int_{t}^{T} r(s)ds}|\mathcal{F}_{t}]\\
                      &=\mathbb{E^Q}[S_{T}^{n} e^{-\int_{t}^{T} r(s)ds} I_{\{S_{T}^{n}>K^{n}\}} |\mathcal{F}_{t}]-K^{n}\mathbb{E^Q}[e^{-\int_{t}^{T} r(s)ds}I_{\{S_{T}^{n}>K^{n}\}}|\mathcal{F}_{t}]\\
                      &\stackrel{\bigtriangleup}{=} A-B
\end{array}
\nonumber
\end{equation}

Since,
\begin{equation}
\rho_t^{'}=\frac{Cov(\frac{\tilde{X}_Q}{\sigma_{\tilde{X}_Q}},\frac{\tilde{Y}_Q}{\sigma_{\tilde{Y}_Q}})}{ \sqrt{D(\frac{\tilde{X}_Q}{\sigma_{\tilde{X}_Q}})}\sqrt{D(\frac{\tilde{Y}_Q}{\sigma_{\tilde{Y}_Q}}})}
         =\frac{ {Cov(\tilde{X}_Q,\tilde{Y}_Q)}}{ \sigma_{\tilde{X}_Q}\sigma_{\tilde{Y}_Q}}= \frac{ \int_{t}^{T}   \rho_u \sigma_{r}(u)\sigma_{S}(u)m(u,T)du}{ \sigma_{\tilde{X}_Q}\sigma_{\tilde{Y}_Q}},
\nonumber
\end{equation}
combining formulae (24)(25)(26), Lemma 3 and Lemma 5, we obtain

\begin{equation}
\begin{array}{ll}
A&= \mathbb{E^Q}[S_{t}^{n} e^{(n-1)\int_{t}^{T} r(s)ds -n\int_{t}^{T} q(s)ds -\int_{t}^{T}\frac{n}{2} \sigma_{S}^{2}(s)ds + \int_{t}^{T} n\sigma_{S}(s)dB^Q(s)} I_{\{\int_{t}^{T} (r(s)-q(s)-\frac{1}{2} \sigma_{S}^{2}(s))ds + \int_{t}^{T} \sigma_{S}(s)dB^Q(s)>\ln\frac{K}{S_{t}}\}} |\mathcal{F}_{t}]\\
 &= \mathbb{E^Q}[S_{t}^{n} e^{(n-1)(G(t,T,r_{t})+\tilde{X}_Q)-n\int_{t}^{T} q(s)ds -\int_{t}^{T}\frac{n}{2} \sigma_{S}^{2}(s)ds + n\tilde{Y}_Q } I_{\{G(t,T,r_{t})+\tilde{X}_Q-\int_{t}^{T} q(s)ds-\int_{t}^{T} \frac{1}{2} \sigma_{S}^{2}(s)ds+\tilde{Y}_Q>\ln\frac{K}{S_{t}}\}} |\mathcal{F}_{t}]\\
 &= S_{t}^{n} e^{(n-1)G(t,T,r_{t})-n\int_{t}^{T} q(s)ds -\int_{t}^{T}\frac{n}{2} \sigma_{S}^{2}(s)ds }
 \mathbb{E^Q}[e^{(n-1)\tilde{X}_Q+ n\tilde{Y}_Q} I_{\{\tilde{X}_Q+\tilde{Y}_Q>\ln\frac{K}{S_{t}}-G(t,T,r_{t})+\int_{t}^{T} q(s)ds+\int_{t}^{T}\frac{1}{2} \sigma_{S}^{2}(s)ds \}} |\mathcal{F}_{t}] \\
 &= S_{t}^{n} e^{(n-1)G(t,T,r_{t})-n\int_{t}^{T} q(s)ds -\frac{n}{2} \sigma_{\tilde{Y}_Q}^{2}} \mathbb{E}[e^{(n-1)\tilde{X}_Q+ n\tilde{Y}_Q} I_{\{\tilde{X}_Q+\tilde{Y}_Q>\ln\frac{K}{S_{t}}-G(t,T,r_{t})+\int_{t}^{T} q(s)ds+\frac{1}{2} \sigma_{\tilde{Y}_Q}^{2}\}} |\mathcal{F}_{t}]\\
 &= S_{t}^{n} e^{(n-1)G(t,T,r_{t})-n\int_{t}^{T} q(s)ds -\frac{n}{2} \sigma_{\tilde{Y}_Q}^{2}} e^{\frac{1}{2}[(n-1)^2 \sigma_{\tilde{X}_Q}^{2}+n^2 \sigma_{\tilde{Y}_Q}^{2}+2\rho_t^{'}n(n-1)\sigma_{\tilde{X}_Q}\sigma_{\tilde{Y}_Q}]}\\
 & \quad  N\Big( \frac{\ln\frac{S_{t}}{K}+G(t,T,r_{t})-\int_{t}^{T} q(s)ds+(n-1)\sigma_{\tilde{X}_Q}^{2}+(n-\frac{1}{2})\sigma_{\tilde{Y}_Q}^{2}+\rho_t^{'}(2n-1)\sigma_{\tilde{X}_Q}\sigma_{\tilde{Y}_Q}} { \sqrt{\sigma_{\tilde{X}_Q}^{2}+\sigma_{\tilde{Y}_Q}^{2}+2\rho_t^{'}\sigma_{\tilde{X}_Q}\sigma_{\tilde{Y}_Q}}}\Big)\\
 &= S_{t}^{n} e^{(n-1)G(t,T,r_{t})-n\int_{t}^{T} q(s)ds +\frac{(n-1)^2}{2}\sigma_{\tilde{X}_Q}^{2}+\frac{n(n-1)}{2} \sigma_{\tilde{Y}_Q}^{2}+\rho_t^{'}n(n-1)\sigma_{\tilde{X}_Q}\sigma_{\tilde{Y}_Q}}\\
 & \quad  N\Big(\frac{\ln\frac{S_{t}}{K}+G(t,T,r_{t})-\int_{t}^{T} q(s)ds +(n-1)\sigma_{\tilde{X}_Q}^{2}+(n-\frac{1}{2})\sigma_{\tilde{Y}_Q}^{2}+\rho_t^{'}(2n-1)\sigma_{\tilde{X}_Q}\sigma_{\tilde{Y}_Q}} { \sqrt{\sigma_{\tilde{X}_Q}^{2}+\sigma_{\tilde{Y}_Q}^{2}+2\rho_t^{'}\sigma_{\tilde{X}_Q}\sigma_{\tilde{Y}_Q}}}\Big)\\
\end{array}
\nonumber
\end{equation}

\begin{equation}
\begin{array}{ll}
B&= K^{n}\mathbb{E^Q}[e^{-G(t,T,r_{t})-\tilde{X}_Q}I_{\{\tilde{X}_Q+\tilde{Y}_Q>\ln\frac{K}{S_{t}}-G(t,T,r_{t})+\int_{t}^{T} q(s)ds+ \frac{1}{2} \sigma_{\tilde{Y}_Q}^{2}\}} |\mathcal{F}_{t}]\\
 &=K^{n} e^{-G(t,T,r_{t})}\mathbb{E}[e^{-\tilde{X}_Q}I_{\{\tilde{X}_Q+\tilde{Y}_Q>\ln\frac{K}{S_{t}}-G(t,T,r_{t})+\int_{t}^{T} q(s)ds+ \frac{1}{2} \sigma_{\tilde{Y}_Q}^{2}\}} |\mathcal{F}_{t}]\\
 &= K^{n} e^{ -G(t,T,r_{t})+ \frac{1}{2}\sigma_{\tilde{X}_Q}^{2} } N\Big( \frac{\ln\frac{S_{t}}{K}+G(t,T,r_{t})-\int_{t}^{T} q(s)ds-(\sigma_{\tilde{X}_Q}^{2}+\frac{\sigma_{\tilde{Y}_Q}^{2}}{2}+\rho_t^{'}\sigma_{\tilde{X}_Q}\sigma_{\tilde{Y}_Q})} {\sqrt{\sigma_{\tilde{X}_Q}^{2}+\sigma_{\tilde{Y}_Q}^{2}+2\rho_t^{'}\sigma_{\tilde{X}_Q}\sigma_{\tilde{Y}_Q}}}\Big)
\end{array}
\nonumber
\end{equation}
Hence formula (27) is proved.
Similarly,  formula (28) can be derived. According to the property of normal distribution $N(-x)=1-N(x)$, formula (29) can be easily verified. Therefore, it completes the proof. \\
\end{proof}

{\bf Theorem 4. } Suppose $Z(t)$ and $B(t)$  are two standard one-dimension Brown motions on probability space with correlation coefficient $\rho_t$ $( -1\leq\rho_t \leq 1)$,  interest rate and stock process satisfy stochastic differential equations (19)(21)(24), with strike price $K$, and maturity time $T$, the European power call option price  $V_{c}(t,r_{t},S_{t},T,K)$ at time $t$ $(0\leq t\leq T)$ with payoff function (3) is

\begin{equation}
\begin{array}{ll}
V_{c}(t,r_{t},S_{t},T,K)= S_{t}^{n} e^{ (n-1)G(t,T,r_{t})-n\int_{t}^{T} q(s)ds +\frac{(n-1)^2}{2}\sigma_{\tilde{X}_Q}^{2}+\frac{n(n-1)}{2} \sigma_{\tilde{Y}_Q}^{2}+\rho_t^{'}n(n-1)\sigma_{\tilde{X}_Q}\sigma_{\tilde{Y}_Q}}N(d_{1})\\
                  \hspace*{2.0cm}  -K e^{ -G(t,T,r_{t})+\frac{\sigma_{\tilde{X}_Q}^{2}}{2}}N(d_{2})
\end{array}
\end{equation}

European power put option $V_{p}(t,r_{t},S_{t},T,K)$ at time $t (0\leq t\leq T)$ with payoff function (5) is

\begin{equation}
\begin{array}{ll}
V_{p}(t,r_{t},S_{t},T,K)= K e^{ -G(t,T,r_{t})+\frac{\sigma_{\tilde{X}_Q}^{2}}{2}}N(-d_{2})\\
                         \hspace*{2.0cm} -S_{t}^{n} e^{ (n-1)G(t,T,r_{t})-n\int_{t}^{T} q(s)ds +\frac{(n-1)^2}{2}\sigma_{\tilde{X}_{Q}}^{2}+\frac{n(n-1)}{2} \sigma_{\tilde{Y}_{Q}}^{2}+\rho_t^{'}n(n-1)\sigma_{\tilde{X}_{Q}}\sigma_{\tilde{Y}_{Q}}} N(-d_{1})
\end{array}
\end{equation}

European power option put--call parity is
\begin{equation}
    \begin{array}{ll}
    &V_{c}(t,r_{t},S_{t},T,K)+K e^{ -G(t,T,r_{t})+\frac{\sigma_{\tilde{X}_{Q}}^{2}}{2}}\\
    &=V_{p}(t,r_{t},S_{t},T,K)+S_{t}^{n} e^{ (n-1)G(t,T,r_{t})-n\int_{t}^{T} q(s)ds +\frac{(n-1)^2}{2}\sigma_{\tilde{X}_{Q}}^{2}+\frac{n(n-1)}{2} \sigma_{\tilde{Y}_{Q}}^{2}+\rho_t^{'}n(n-1)\sigma_{\tilde{X}_{Q}}\sigma_{\tilde{Y}_{Q}}}
    \end{array}
    \end{equation}

where,
 \begin{equation}
     \begin{array}{ll}
     d_{1}=\frac{\ln\frac{S_{t}}{\sqrt[n]{K}}+G(t,T,r_{t})-\int_{t}^{T} q(s)ds+(n-1)\sigma_{\tilde{X}_{Q}}^{2}+(n-\frac{1}{2})\sigma_{\tilde{Y}_{Q}}^{2}+\rho_t^{'}(2n-1)\sigma_{\tilde{X}_{Q}}\sigma_{\tilde{Y}_{Q}}} {\sqrt{\sigma_{\tilde{X}_{Q}}^{2}+\sigma_{\tilde{Y}_{Q}}^{2}+2\rho_t^{'}\sigma_{\tilde{X}_{Q}}\sigma_{\tilde{Y}_{Q}}}},\\
     d_{2}=\frac{\ln\frac{S_{t}}{\sqrt[n]{K}}+G(t,T,r_{t})-\int_{t}^{T} q(s)ds-(\sigma_{\tilde{X}_{Q}}^{2}+\frac{\sigma_{\tilde{Y}_{Q}}^{2}}{2}+\rho_t^{'}\sigma_{\tilde{X}_{Q}}\sigma_{\tilde{Y}_{Q}})}
     {\sqrt{\sigma_{\tilde{X}_{Q}}^{2}+\sigma_{\tilde{Y}_{Q}}^{2}+2\rho_t^{'}\sigma_{\tilde{X}_{Q}}\sigma_{\tilde{Y}_{Q}}}},\\
     \rho_t^{'}= \frac{ \int_{t}^{T} \rho_u \sigma_{r}(u)\sigma_{u}m(u,T)du}{\sigma_{\tilde{X}_{Q}}\sigma_{\tilde{Y}_{Q}}}
     \end{array}
     \end{equation}
 where $N(x)$ denotes standard normal distribution. \\

\begin{proof} According to arbitrage risk-neutral theory, the price of  European power call option with payoff function (3) at time $t$ is
\begin{equation}
\begin{array}{ll}
V_{c}(t,r_{t},S_{t},T,K)&=\mathbb{E^Q}[ (S_{T}^{n}-K)^{+} e^{-\int_{t}^{T} r(s)ds}|\mathcal{F}_{t}]\\
                      &=\mathbb{E^Q}[S_{T}^{n} e^{-\int_{t}^{T} r(s)ds}I_{\{S_{T}^{n}>K\}} |\mathcal{F}_{t}]-K \mathbb{E^Q}[e^{-\int_{t}^{T} r(s)ds}I_{\{S_{T}^{n}>K\}}|\mathcal{F}_{t}]\\
                      &\stackrel{\bigtriangleup}{=} A-B
\end{array}
\nonumber
\end{equation}

Since,
\begin{equation}
\rho_t^{'}=\frac{  Cov(\frac{\tilde{X}_Q}{\sigma_{\tilde{X}_Q}},\frac{\tilde{Y}_Q}{\sigma_{\tilde{Y}_Q}})}{\sqrt{D(\frac{\tilde{X}_Q}{\sigma_{\tilde{X}_Q}})}\sqrt{D(\frac{\tilde{Y}_Q}{\sigma_{\tilde{Y}_Q}}})}
         =\frac{{Cov(\tilde{X}_Q,\tilde{Y}_Q)}}{ \sigma_{\tilde{X}_Q}\sigma_{\tilde{Y}_Q}}
         =\frac{\int_{t}^{T} \rho_u\sigma_{r}(u)\sigma_{u}m(u,T)du}{\sigma_{\tilde{X}_Q}\sigma_{\tilde{Y}_Q}},
\nonumber
\end{equation}
combining formulae (24)(25)(26), Lemma 3 and Lemma 5, we obtain
\begin{equation}
\begin{array}{ll}
A&=\mathbb{E^Q}[S_{t}^{n} e^{(n-1)\int_{t}^{T} r(s)ds -n\int_{t}^{T} q(s)ds-\int_{t}^{T}\frac{n}{2} \sigma_{S}^{2}(s)ds + \int_{t}^{T} n\sigma_{S}(s)dB_{s}}I_{\{\int_{t}^{T} (r(s)-q(s)-\frac{1}{2} \sigma_{S}^{2}(s))ds + \int_{t}^{T} \sigma_{S}(s)dB_{s}>\ln\frac{\sqrt[n]{K}}{S_{t}}\}} |\mathcal{F}_{t}]\\
 &=\mathbb{E^Q}[S_{t}^{n} e^{(n-1)(G(t,T,r_{t})+\tilde{X}_Q)-n\int_{t}^{T} q(s)ds -\int_{t}^{T}\frac{n}{2} \sigma_{S}^{2}(s)ds + n\tilde{Y}_Q } I_{\{G(t,T,r_{t})-\int_{t}^{T} q(s)ds+\tilde{X}_Q-\int_{t}^{T}\frac{1}{2} \sigma_{S}^{2}(s)ds+\tilde{Y}_Q>\ln\frac{\sqrt[n]{K}}{S_{t}}\}} |\mathcal{F}_{t}]\\
 &= S_{t}^{n} e^{(n-1)G(t,T,r_{t})-n\int_{t}^{T} q(s)ds -\int_{t}^{T}\frac{n}{2} \sigma_{S}^{2}(s)ds } \mathbb{E^Q}[e^{(n-1)\tilde{X}_Q+ n\tilde{Y}_Q}
 I_{\{\tilde{X}_Q+\tilde{Y}_Q>\ln\frac{\sqrt[n]{K}}{S_{t}}-G(t,T,r_{t})+\int_{t}^{T} q(s)ds+\int_{t}^{T}\frac{1}{2} \sigma_{S}^{2}(s)ds \}} |\mathcal{F}_{t}]\\
 &= S_{t}^{n} e^{(n-1)G(t,T,r_{t})-n\int_{t}^{T} q(s)ds -\frac{n}{2} \sigma_{\tilde{Y}_Q}^{2}} \mathbb{E^Q}[e^{(n-1)\tilde{X}_Q+ n\tilde{Y}_Q}I_{\{\tilde{X}_Q+\tilde{Y}_Q>\ln\frac{\sqrt[n]{K}}{S_{t}}-G(t,T,r_{t})+\int_{t}^{T} q(s)ds+\frac{1}{2} \sigma_{\tilde{Y}_Q}^{2}\}} |\mathcal{F}_{t}]\\
 &= S_{t}^{n} e^{(n-1)G(t,T,r_{t}) -n\int_{t}^{T} q(s)ds-\frac{n}{2} \sigma_{\tilde{Y}_Q}^{2}} e^{\frac{1}{2}[(n-1)^2 \sigma_{\tilde{X}_Q}^{2}+n^2 \sigma_{\tilde{Y}_Q}^{2}+2\rho_t^{'}n(n-1)\sigma_{\tilde{X}_Q}\sigma_{\tilde{Y}_Q}]}\\
 & \quad  N\Big(\frac{\ln\frac{S_{t}}{\sqrt[n]{K}}+G(t,T,r_{t})-\int_{t}^{T} q(s)ds+(n-1)\sigma_{\tilde{X}_Q}^{2}+(n-\frac{1}{2})\sigma_{\tilde{Y}_Q}^{2}+\rho_t^{'}(2n-1)\sigma_{\tilde{X}_Q}\sigma_{\tilde{Y}_Q}} { \sqrt{\sigma_{\tilde{X}_Q}^{2}+\sigma_{\tilde{Y}_Q}^{2}+2\rho_t^{'}\sigma_{\tilde{X}_Q}\sigma_{\tilde{Y}_Q}}}\Big)\\
 &= S_{t}^{n} e^{(n-1)G(t,T,r_{t})-n\int_{t}^{T} q(s)ds +\frac{(n-1)^2}{2}\sigma_{\tilde{X}_Q}^{2}+\frac{n(n-1)}{2} \sigma_{\tilde{Y}_Q}^{2}+\rho_t^{'}n(n-1)\sigma_{\tilde{X}_Q}\sigma_{\tilde{Y}_Q}}\\
 & \quad  N\Big(\frac{\ln\frac{S_{t}}{\sqrt[n]{K}}+G(t,T,r_{t})-\int_{t}^{T} q(s)ds +(n-1)\sigma_{\tilde{X}_Q}^{2}+(n-\frac{1}{2})\sigma_{\tilde{Y}_Q}^{2}+\rho_t^{'}(2n-1)\sigma_{\tilde{X}_Q}\sigma_{\tilde{Y}_Q}} { \sqrt{\sigma_{\tilde{X}_Q}^{2}+\sigma_{\tilde{Y}_Q}^{2}+2\rho_t^{'}\sigma_{\tilde{X}_Q}\sigma_{\tilde{Y}_Q}}}\Big)\\
\end{array}
\nonumber
\end{equation}

\begin{equation}
\begin{array}{ll}
B&= K\mathbb{E^Q}[e^{-G(t,T,r_{t})-\tilde{X}_Q}I_{\{\tilde{X}_Q+\tilde{Y}_Q>\ln\frac{\sqrt[n]{K}}{S_{t}}-G(t,T,r_{t})+\int_{t}^{T} q(s)ds+ \frac{1}{2} \sigma_{\tilde{Y}_Q}^{2}\}} |\mathcal{F}_{t}]\\
 &=K e^{-G(t,T,r_{t})}\mathbb{E^Q}[e^{-\tilde{X}_Q}I_{\{\tilde{X}_Q+\tilde{Y}_Q>\ln\frac{\sqrt[n]{K}}{S_{t}}-G(t,T,r_{t})+\int_{t}^{T} q(s)ds+ \frac{1}{2} \sigma_{\tilde{Y}_Q}^{2}\}} |\mathcal{F}_{t}]\\
 &= K e^{-G(t,T,r_{t})+\frac{1}{2}\sigma_{\tilde{X}_Q}^{2} } N\Big(\frac{\ln\frac{S_{t}}{\sqrt[n]{K}}+G(t,T,r_{t})-\int_{t}^{T} q(s)ds-(\sigma_{\tilde{X}_Q}^{2}+\frac{\sigma_{\tilde{Y}_Q}^{2}}{2}+\rho_t^{'}\sigma_{\tilde{X}_Q}\sigma_{\tilde{Y}_Q})} {\sqrt{\sigma_{\tilde{X}_Q}^{2}+\sigma_{\tilde{Y}_Q}^{2}+2\rho_t^{'}\sigma_{\tilde{X}_Q}\sigma_{\tilde{Y}_Q}}}\Big)
\end{array}
\nonumber
\end{equation}

Hence, formulae (31) is proved. And, Formula (32) can be derived similarly. According to the property of normal distribution $N(-x)=1-N(x)$, formula (33) can be easily verified. Therefore, it completes the proof. \\
\end{proof}

{\bf Remark } Though  Theorem 1 and Theorem 3 take almost uniform, so do Theorem 2 and Theorem 4, their notations represent different meaning.

\section{Vasic\v{e}k interest rate is under equivalent martingale measurement with numeraire change}

In fact, {\bf Market Assumption 1} can be taken the following concise form:

\begin{equation}
\left\{\begin{array}{l}
      \displaystyle dS(t)=(r(t)-q(t)) S(t) dt +\sigma_{S}(t) S(t) dB^Q(t)\\
      \displaystyle d r_{t}= \displaystyle (\beta_{t}-\alpha_{t}r_{t})dt+  \sigma_{r}(t)d Z^Q(t)
\end{array}\right.
\end{equation}
where $dB^Q(t) d Z^Q(t)=\rho_t dt $.

 As in Shreve,2004, we denote discount process  $D(t)=\exp\{-\int_{0}^{t} r_s ds\}$, the price of a zero-coupon bond paying 1 at time $T$ under risk--neutral probability measurement $Q$ is
\begin{equation}
\begin{array}{l}
\displaystyle P(t,T)=\frac{1}{D(t)} \mathbb{E^Q} [D(T)|\mathcal{F}_{t}]\\
\end{array}
\end{equation}
where $P(T,T)=1$.

As in Li, et al. (2008), calculating $P(t,T)$, we obtain

 \begin{equation}
\begin{array}{ll}
\displaystyle P(t,T)&=\frac{1}{D(t)} \mathbb{E^Q} [D(T)|\mathcal{F}_{t}]=\mathbb{E^Q} [e^{-\int_{t}^{T} r_s ds}|\mathcal{F}_{t}]\\
         &=\exp\{-r_t m(t,T)-\int_{t}^{T} \beta_s m(s,T)ds +\frac{1}{2} \int_{t}^{T} \sigma_r^2(s) m^2(s,T) ds  \} \\
\end{array}
\end{equation}

Applying Ito formula to above formula, we obtain

\begin{equation}
\begin{array}{ll}
\displaystyle \frac{d P(t,T)}{ P(t,T)}=r_t dt-\sigma_r(t) m(t,T) dZ^Q(t) \\
\end{array}
\end{equation}

{\bf Lemma 7.} For $P(t,T)$ with $r(t)$ in formula (38), $D(t)P(t,T)$ is a $Q$-martingale.

\begin{proof} Applying Ito formula,
\begin{equation}
\begin{array}{l}
\displaystyle d(D(t)P(t,T))=-D(t)P(t,T)\sigma_r(t)m(t,T) dZ^Q(t)=D(t)P(t,T)\sigma_r(t)m(t,T) d(-Z^Q(t)).\\
\end{array}
\end{equation}
Hence, it is a $Q$-martingale.
\end{proof}

Applying Ito formula, we obtain

{\bf Lemma 8.}
\begin{equation}
\begin{array}{ll}
\displaystyle d\frac{S_t}{P(t,T)}=\displaystyle \frac{S_t}{P(t,T)} [(-q(t)-\sigma_r(t)m(t,T))dt+ \sigma_S(t)dB_t^Q+ \sigma_r(t)m(t,T) dZ^Q(t)].\\
\displaystyle d\ln\frac{S_t}{P(t,T)}=\displaystyle [(-q(t)-\frac{1}{2}\sigma_S^2(t)+\frac{1}{2}\sigma_r^2(t)m^2(t,T))dt+ \sigma_S(t)dB_t^Q+ \sigma_r(t)m(t,T) dZ^Q(t)].\\
\displaystyle \frac{S_T}{P(T,T)}=\displaystyle \frac{S_t}{P(t,T)}e^{ \{\int_{t}^{T}(-q(s)-\frac{1}{2}\sigma_S^2(s)+\frac{1}{2}\sigma_r^2(s)m^2(s,T))ds
 +\int_{t}^{T} \sigma_S(s)dB_s^Q+ \int_{t}^{T} \sigma_r(s)m(s,T) dZ^Q(s)\} }.\\
\end{array}
\end{equation}

Since, $dB^Q(t) d Z^Q(t)=\rho_t dt $, there exists $B^{\perp,Q}(t)$ under $Q$ that $B^{Q}(t)$ and $B^{\perp,Q}(t)$ are independent Brownian motions and
\begin{equation}
d Z^Q(t)=\rho_t d B^Q(t) + \sqrt{1-\rho^2_t} d B^{\perp,Q}(t).
\end{equation}

\begin{equation}
\begin{array}{ll}
\displaystyle \frac{d P(t,T)}{ P(t,T)}=r_t dt- [\rho_t \sigma_r(t) m(t,T) dB^Q(t)+ \sqrt{1-\rho^2_t} \sigma_r(t) m(t,T) d B^{\perp,Q}(t)]
\end{array}
\end{equation}
Take T-forward measure with zero-coupon bond $P(t,T)$ as a numeraire, the T--forward measure $P^T$ is defined by
\begin{equation}
\begin{array}{l}
\displaystyle P^T(A)= \frac{1}{P(0,T)} \int_{A} D(T) d Q, \quad  for \quad all \quad A \in \mathcal{F}.\\
\end{array}
\end{equation}
Let
\begin{equation}
\begin{array}{l}
\displaystyle B^{P^T}(t)=B^Q(t)+\int_{0}^{t} \rho_s \sigma_r(s)m(s,T) ds, \quad B^{\perp,P^T}(t)=B^{\perp,Q}(t)+\int_{0}^{t} \sqrt{1-\rho^2_s}\sigma_r(s)m(s,T) ds .\\
\end{array}
\end{equation}

Then, according to Girsanov Theorem, $B^{P^T}(t)$ and $B^{\perp,P^T}(t)$ are independent Brownian motions under $P^T$.

Denote
\begin{equation}
\begin{array}{ll}
X_{P^T}=\int_{t}^{T} \rho_u \sigma_{r}(u) m(u,T)dB^{P^T}(u)+\int_{t}^{T}  \sqrt{1-\rho_u^2} \sigma_{r}(u)m(u,T)dB^{\perp,{P^T}}(u)\\
Y_{P^T}=\int_{t}^{T} \sigma_{S}(u)dB^{P^T}(u),\\
\end{array}
\end{equation}

We obtain
\begin{equation}
\begin{array}{ll}
\sigma_{X_{P^T}}^{2}=\int_{t}^{T} \sigma_{r}^{2}(u)m^{2}(u,T)du,  \quad \sigma_{Y_{P^T}}^{2}=\int_{t}^{T} \sigma_{S}^{2}(u)du    \\
\end{array}
\end{equation}
And,
\begin{equation}
\begin{array}{ll}
\displaystyle \frac{S(T)}{P(T,T)}=\displaystyle \frac{S(t)}{P(t,T)} e^{\int_{t}^{T}(-q(s)-\frac{1}{2}\sigma_S^2(s)-\frac{1}{2}\sigma_r^2(s)m^2(s,T)-\rho_s \sigma_S(s)\sigma_r(s)m(s,T))ds +X_{P^T}+ Y_{P^T}}.\\
\end{array}
\end{equation}

Thus, we obtain Radon-Nikodym derivatives of probability measure $ P^T $ with respect to $Q$,  the price of  payoff $V(T)$ at time t is
\begin{equation}
\begin{array}{ll}
\displaystyle V(t,r_{t},S_{t},P(t,T),T,K)=\displaystyle \frac{1}{D(t)} \mathbb{E^Q} [D(T)V(T)|\mathcal{F}_{t}] \\
     =\displaystyle \frac{1}{D(t)} \mathbb{E^Q} [D(T)P(T,T)V(T)|\mathcal{F}_{t}] \\
     =\displaystyle P(t,T) \mathbb{E}^{P^T} [V(T)|\mathcal{F}_{t}] \\
\end{array}
\end{equation}

{\bf Theorem 5.} Suppose $Z(t)$ and $B(t)$ are two standard one-dimension Brown motions on probability space with correlation coefficient $\rho_t$ $( -1\leq\rho_t \leq 1)$ of exponential Ornstein-Uhlenbeck stock process (1) and Vasicek interest rate (7) and stochastic differential equations (35), with strike price $K$, and maturity time $T$, European power call option price $V_{c}(t,r_{t},S_{t},P(t,T),T,K)$ at time $t (0\leq t\leq T)$ with payoff function (2) is

\begin{equation}
\begin{array}{ll}
V_{c}(t,r_{t},S_{t},P(t,T),T,K)=\frac{S^n_t}{P^{n-1}(t,T)} e^{\frac{n(n-1)}{2}\sigma_{X_{P^T}}^{2}+\frac{n(n-1)}{2} \sigma_{Y_{P^T}}^{2}+\rho_t^{'}n(n-1)\sigma_{X_{P^T}}\sigma_{Y_{P^T}}-n\int_{t}^{T} q(s)ds  }N( d_1)\\
\hspace*{2.0cm}-P(t,T) K^{n}N(d_2) \\
\end{array}
\end{equation}

European power put option price $V_{p}(t,r_{t},S_{t},P(t,T),T,K)$ at time $t (0\leq t\leq T)$ with payoff function (4) is

     \begin{equation}
     \begin{array}{ll}
      V_{p}(t,r_{t},S_{t},P(t,T),T,K)=P(t,T) K^{n}N(-d_2)\\
                         \hspace*{2.0cm} -\frac{S^n_t}{P^{n-1}(t,T)} e^{\frac{n(n-1)}{2}\sigma_{X_{P^T}}^{2}+\frac{n(n-1)}{2} \sigma_{Y_{P^T}}^{2}+\rho_t^{'}n(n-1)\sigma_{X_{P^T}}\sigma_{Y_{P^T}}-n\int_{t}^{T} q(s)ds  } N(-d_1)\\
     \end{array}
     \end{equation}

European power put--call parity is

    \begin{equation}
    \begin{array}{ll}
    &V_{c}(t,r_{t},S_{t},P(t,T),T,K)+P(t,T) K^{n}\\
    &=V_{p}(t,r_{t},S_{t},P(t,T),T,K)+\frac{S^n_t}{P^{n-1}(t,T)} e^{\frac{n(n-1)}{2}\sigma_{X_{P^T}}^{2}+\frac{n(n-1)}{2} \sigma_{Y_{P^T}}^{2}+\rho_t^{'}n(n-1)\sigma_{X_{P^T}}\sigma_{Y_{P^T}}-n\int_{t}^{T} q(s)ds  }
     \end{array}
    \end{equation}
where,
     \begin{equation}
     \begin{array}{ll}
     d_{1}=\frac{\ln\frac{S_{t}}{KP(t,T)}-\int_{t}^{T} q(s)ds +(n-\frac{1}{2})\sigma_{X_{P^T}}^{2}+(n-\frac{1}{2})\sigma_{Y_{P^T}}^{2}+(2n-1)\rho_t^{'}\sigma_{X_{P^T}}\sigma_{Y_{P^T}}} { \sqrt{\sigma_{X_{P^T}}^{2}+\sigma_{Y_{P^T}}^{2}+2\rho_t^{'}\sigma_{X_{P^T}}\sigma_{Y_{P^T}}}},\\
     d_{2}=\frac{\ln \frac{S_t}{KP(t,T)}-\int_{t}^{T}q(s)ds-\frac{1}{2}\sigma_{ Y_{P^T}}^2- \frac{1}{2}\sigma_{X_{P^T}}^2-\rho^{'}_t \sigma_{ X_{P^T}} \sigma_{ Y_{P^T}}} {\sqrt{\sigma_{X_{P^T}}^{2}+\sigma_{Y_{P^T}}^{2}+2\rho_t^{'}\sigma_{X_{P^T}}\sigma_{Y_{P^T}}}}\\
     \rho_t^{'}= \frac{ \int_{t}^{T} \rho_u \sigma_{r}(u)\sigma_{u}m(u,T)du}{ \sigma_{X_{P^T}}\sigma_{Y_{P^T}}}
     \end{array}
     \end{equation}

where $N(x)$ denotes standard normal distribution.\\

\begin{proof}  According to arbitrage risk-neutral theory, using T-forward measure with zero-coupon bond as a numeraire, European power call option at time $t$  with  payoff function $\xi=(S_{T}^{n}-K^{n})^{+}$ is
\begin{equation}
\begin{array}{ll}
V_{c}(t,r_{t},S_{t},P(t,T),T,K)&=\mathbb{E^Q}[ (S_{T}^{n}-K^{n})^{+} e^{-\int_{t}^{T} r(s)ds}|\mathcal{F}_{t}]\\
                      &=P(t,T)\mathbb{E}^{P^T} [ (S_{T}^{n}-K^{n})^{+}|\mathcal{F}_{t}]\\
                      &=P(t,T)\mathbb{E}^{P^T} [S_{T}^{n} I_{\{S_{T}^{n}>K^{n}\}} |\mathcal{F}_{t}]
                        -P(t,T) K^{n}\mathbb{E}^{P^T}[I_{\{S_{T}^{n}>K^{n}\}}|\mathcal{F}_{t}]\\
                      &=P(t,T)\mathbb{E}^{P^T} [S_{T}^{n} I_{\{S_{T}>K\}} |\mathcal{F}_{t}]
                        -P(t,T) K^{n} {P^T}[S_{T}>K|\mathcal{F}_{t}]\\
                      &=P(t,T)\mathbb{E}^{P^T} [(\frac{S_{T}}{P(T,T)})^{n} I_{\{\frac{S_{T}}{P(T,T)}>K\}} |\mathcal{F}_{t}]
                        -P(t,T) K^{n} {P^T}[\frac{S_{T}}{P(T,T)}>K|\mathcal{F}_{t}]\\
                       &\stackrel{\bigtriangleup}{=} A-B
\end{array}
\nonumber
\end{equation}

Since,
\begin{equation}
\rho_t^{'}=\frac{Cov(\frac{X_{P^T}}{\sigma_{X_{P^T}}},\frac{Y_{P^T}}{\sigma_{Y_{P^T}}})}{ \sqrt{D(\frac{X_{P^T}}{\sigma_{X_{P^T}}})}\sqrt{D(\frac{Y_{P^T}}{\sigma_{Y_{P^T}}}})}
         =\frac{ {Cov(X_{P^T},Y_{P^T})}}{ \sigma_{X_{P^T}}\sigma_{Y_{P^T}}}
         = \frac{ \int_{t}^{T} \rho_u \sigma_{r}(u)\sigma_{S}(u)m(u,T)du}{ \sigma_{X_{P^T}}\sigma_{Y_{P^T}}},
\nonumber
\end{equation}
combining formulae (47)(48) and Lemma 3, we obtain

\begin{equation}
\begin{array}{ll}
A&=P(t,T)\mathbb{E}^{P^T} [(\frac{S_t}{P(t,T)})^n e^{n [\int_{t}^{T}(-q(s)-\frac{1}{2}\sigma_S^2(s)-\frac{1}{2}\sigma_r^2(s)m^2(s,T)-\rho_s \sigma_S(s)\sigma_r(s)m(s,T))ds+X_{P^T}+ Y_{P^T}]} \\
&\quad I_{\{\frac{S_t}{P(t,T)} e^{\int_{t}^{T}(-q(s)-\frac{1}{2}\sigma_S^2(s)-\frac{1}{2}\sigma_r^2(s)m^2(s,T)-\rho_s \sigma_S(s)\sigma_r(s)m(s,T))ds+X_{P^T}+ Y_{P^T}}>K\}} |\mathcal{F}_{t}]\\
&=P(t,T)\mathbb{E}^{P^T} [(\frac{S_t}{P(t,T)})^n e^{n [-\int_{t}^{T}q(s)ds -\frac{1}{2}\sigma_{ Y_{P^T}}^2-\frac{1}{2}\sigma_{X_{P^T}}^2-\rho^{'}_t \sigma_{ X_{P^T}} \sigma_{ Y_{P^T}} + X_{P^T}+ Y_{P^T}]} \\
&\quad I_{\{\frac{S_t}{P(t,T)} e^{-\int_{t}^{T}q(s) ds -\frac{1}{2}\sigma_{ Y_{P^T}}^2-\frac{1}{2}\sigma_{X_{P^T}}^2-\rho^{'}_t \sigma_{ X_{P^T}} \sigma_{ Y_{P^T}} + X_{P^T}+ Y_{P^T}}>K \}} |\mathcal{F}_{t}]\\
&=P(t,T)(\frac{S_t}{P(t,T)})^n  e^{n [-\int_{t}^{T}q(s) ds -\frac{1}{2}\sigma_{ Y_{P^T}}^2-\frac{1}{2}\sigma_{X_{P^T}}^2-\rho^{'}_t \sigma_{ X_{P^T}} \sigma_{ Y_{P^T}}]}
\mathbb{E}^{P^T} [ e^{n X_{P^T}+ nY_{P^T}} \\
&\quad I_{\{ X_{P^T}+ Y_{P^T}>\ln \frac{KP(t,T)}{S_t}+\int_{t}^{T}q(s)ds+\frac{1}{2}\sigma_{ Y_{P^T}}^2+ \frac{1}{2}\sigma_{X_{P^T}}^2+\rho^{'}_t \sigma_{ X_{P^T}} \sigma_{ Y_{P^T}}\}}|\mathcal{F}_{t}]\\
&=\frac{S^n_t}{P^{n-1}(t,T)} e^{n [-\int_{t}^{T}q(s) ds -\frac{1}{2}\sigma_{ Y_{P^T}}^2-\frac{1}{2}\sigma_{X_{P^T}}^2-\rho^{'}_t \sigma_{ X_{P^T}} \sigma_{ Y_{P^T}}]} \mathbb{E}^{P^T} [ e^{n X_{P^T}+ nY_{P^T}} \\
&\quad I_{\{ X_{P^T}+ Y_{P^T}>\ln \frac{KP(t,T)}{S_t}+\int_{t}^{T}q(s)ds+\frac{1}{2}\sigma_{ Y_{P^T}}^2+ \frac{1}{2}\sigma_{X_{P^T}}^2+\rho^{'}_t \sigma_{ X_{P^T}} \sigma_{Y_{P^T}}\}}|\mathcal{F}_{t}]\\
&= \frac{S^n_t}{P^{n-1}(t,T)} e^{n [-\int_{t}^{T}q(s) ds -\frac{1}{2}\sigma_{ Y_{P^T}}^2-\frac{1}{2}\sigma_{X_{P^T}}^2-\rho^{'}_t \sigma_{ X_{P^T}} \sigma_{ Y_{P^T}}]} e^{\frac{1}{2}[n^2 \sigma_{X_{P^T}}^{2}+n^2 \sigma_{Y_{P^T}}^{2}+2\rho_t^{'}n^2 \sigma_{X_{P^T}}\sigma_{Y_{P^T}}]}\\
& \quad  N\Big( \frac{\ln\frac{S_{t}}{KP(t,T)}-\int_{t}^{T} q(s)ds +(n-\frac{1}{2})\sigma_{X_{P^T}}^{2}+(n-\frac{1}{2})\sigma_{Y_{P^T}}^{2}+(2n-1)\rho_t^{'}\sigma_{X_{P^T}}\sigma_{Y_{P^T}}} { \sqrt{\sigma_{X_{P^T}}^{2}+\sigma_{Y_{P^T}}^{2}+2\rho_t^{'}\sigma_{X_{P^T}}\sigma_{Y_{P^T}}}}\Big)\\
&= \frac{S^n_t}{P^{n-1}(t,T)} e^{\frac{n(n-1)}{2}\sigma_{X_{P^T}}^{2}+\frac{n(n-1)}{2} \sigma_{Y_{P^T}}^{2}+\rho_t^{'}n(n-1)\sigma_{X_{P^T}}\sigma_{Y_{P^T}}-n\int_{t}^{T} q(s)ds  }\\
& \quad  N\Big( \frac{\ln\frac{S_{t}}{KP(t,T)}-\int_{t}^{T} q(s)ds +(n-\frac{1}{2})\sigma_{X_{P^T}}^{2}+(n-\frac{1}{2})\sigma_{Y_{P^T}}^{2}+(2n-1)\rho_t^{'}\sigma_{X_{P^T}}\sigma_{Y_{P^T}}} { \sqrt{\sigma_{X_{P^T}}^{2}+\sigma_{Y_{P^T}}^{2}+2\rho_t^{'}\sigma_{X_{P^T}}\sigma_{Y_{P^T}}}}\Big)\\
&= \frac{S^n_t}{P^{n-1}(t,T)} e^{\frac{n(n-1)}{2}\sigma_{X_{P^T}}^{2}+\frac{n(n-1)}{2} \sigma_{Y_{P^T}}^{2}+\rho_t^{'}n(n-1)\sigma_{X_{P^T}}\sigma_{Y_{P^T}}-n\int_{t}^{T} q(s)ds  } N(d_1)\\
\end{array}
\nonumber
\end{equation}

\begin{equation}
\begin{array}{ll}
B&=P(t,T) K^{n} {P^T}[\frac{S_t}{P(t,T)} e^{\int_{t}^{T}(-q(s)-\frac{1}{2}\sigma_S^2(s)-\frac{1}{2}\sigma_r^2(s)m^2(s,T)-\rho_s \sigma_S(s)\sigma_r(s)m(s,T))ds+X_{P^T}+ Y_{P^T}}>K |\mathcal{F}_{t}]\\
&=P(t,T) K^{n} {P^T}[ X_{P^T}+ Y_{P^T}>\ln \frac{KP(t,T)}{S_t}+\int_{t}^{T}q(s)ds+\frac{1}{2}\sigma_{ Y_{P^T}}^2+ \frac{1}{2}\sigma_{X_{P^T}}^2+\rho^{'}_t \sigma_{ X_{P^T}} \sigma_{ Y_{P^T}} |\mathcal{F}_{t}]\\
&=P(t,T) K^{n} {P^T}[- X_{P^T}- Y_{P^T}<-(\ln \frac{KP(t,T)}{S_t}+\int_{t}^{T}q(s)ds+\frac{1}{2}\sigma_{ Y_{P^T}}^2+ \frac{1}{2}\sigma_{X_{P^T}}^2+\rho^{'}_t \sigma_{ X_{P^T}} \sigma_{ Y_{P^T}}) |\mathcal{F}_{t}]\\
&=P(t,T) K^{n} {P^T}[\frac{- X_{P^T}- Y_{P^T}}{\sqrt{\sigma_{X_{P^T}}^{2}+\sigma_{Y_{P^T}}^{2}+2\rho_t^{'}\sigma_{X_{P^T}}\sigma_{Y_{P^T}}}}<\frac{-(\ln \frac{KP(t,T)}{S_t}+\int_{t}^{T}q(s)ds+\frac{1}{2}\sigma_{ Y_{P^T}}^2+ \frac{1}{2}\sigma_{X_{P^T}}^2+\rho^{'}_t \sigma_{ X_{P^T}} \sigma_{ Y_{P^T}})}{\sqrt{\sigma_{X_{P^T}}^{2}+\sigma_{Y_{P^T}}^{2}+2\rho_t^{'}\sigma_{X_{P^T}}\sigma_{Y_{P^T}}}} |\mathcal{F}_{t}]\\
&=P(t,T) K^{n}N\Big(\frac{\ln \frac{S_t}{KP(t,T)}-\int_{t}^{T}q(s)ds-\frac{1}{2}\sigma_{ Y_{P^T}}^2- \frac{1}{2}\sigma_{X_{P^T}}^2-\rho^{'}_t \sigma_{ X_{P^T}} \sigma_{ Y_{P^T}}} {\sqrt{\sigma_{X_{P^T}}^{2}+\sigma_{Y_{P^T}}^{2}+2\rho_t^{'}\sigma_{X_{P^T}}\sigma_{Y_{P^T}}}}  \Big)\\
&=P(t,T) K^{n}N(d_2)\\
\end{array}
\nonumber
\end{equation}
Hence formula (49) is proved. Similarly,  formula (50) can be derived.
According to the property of normal distribution $N(-x)=1-N(x)$, formula (51) can be easily verified. Therefore, it completes the proof. \\
\end{proof}

{\bf Theorem 6. } Suppose $Z(t)$ and $B(t)$ are two standard one-dimension Brown motions on probability space with correlation coefficient $\rho_t$ $( -1\leq\rho_t \leq 1)$ of exponential Ornstein-Uhlenbeck stock process (1) and Vasicek interest rate (7) and stochastic differential equations (35), with strike price $K$, and maturity time $T$, the European power call option price  $V_{c}(t,r_{t},S_{t},P(t,T),T,K)$ at time $t$ $(0\leq t\leq T)$ with payoff function (3) is

\begin{equation}
\begin{array}{ll}
V_{c}(t,r_{t},S_{t},P(t,T),T,K)=\frac{S^n_t}{P^{n-1}(t,T)} e^{\frac{n(n-1)}{2}\sigma_{X_{P^T}}^{2}+\frac{n(n-1)}{2} \sigma_{Y_{P^T}}^{2}+\rho_t^{'}n(n-1)\sigma_{X_{P^T}}\sigma_{Y_{P^T}}-n\int_{t}^{T} q(s)ds  }N( d_1)\\
\hspace*{2.0cm}-P(t,T) K N(d_2). \\
\end{array}
\end{equation}

European power put option $V_{p}(t,r_{t},S_{t},P(t,T),T,K)$ at time $t (0\leq t\leq T)$ with payoff function (5) is
     \begin{equation}
     \begin{array}{ll}
      V_{p}(t,r_{t},S_{t},P(t,T),T,K)=P(t,T) K N(-d_2)\\
                         \hspace*{2.0cm} -\frac{S^n_t}{P^{n-1}(t,T)} e^{\frac{n(n-1)}{2}\sigma_{X_{P^T}}^{2}+\frac{n(n-1)}{2} \sigma_{Y_{P^T}}^{2}+\rho_t^{'}n(n-1)\sigma_{X_{P^T}}\sigma_{Y_{P^T}}-n\int_{t}^{T} q(s)ds  } N(-d_1).\\
     \end{array}
     \end{equation}

European power option put--call parity is
  \begin{equation}
    \begin{array}{ll}
    &V_{c}(t,r_{t},S_{t},P(t,T),T,K)+P(t,T) K\\
    &=V_{p}(t,r_{t},S_{t},P(t,T),T,K)+\frac{S^n_t}{P^{n-1}(t,T)} e^{\frac{n(n-1)}{2}\sigma_{X_{P^T}}^{2}+\frac{n(n-1)}{2} \sigma_{Y_{P^T}}^{2}+\rho_t^{'}n(n-1)\sigma_{X_{P^T}}\sigma_{Y_{P^T}}-n\int_{t}^{T} q(s)ds  }.
     \end{array}
    \end{equation}
where,
     \begin{equation}
     \begin{array}{ll}
     d_{1}=\frac{\ln\frac{S_{t}}{\sqrt[n]{K} P(t,T)}-\int_{t}^{T} q(s)ds +(n-\frac{1}{2})\sigma_{X_{P^T}}^{2}+(n-\frac{1}{2})\sigma_{Y_{P^T}}^{2}+(2n-1)\rho_t^{'}\sigma_{X_{P^T}}\sigma_{Y_{P^T}}} { \sqrt{\sigma_{X_{P^T}}^{2}+\sigma_{Y_{P^T}}^{2}+2\rho_t^{'}\sigma_{X_{P^T}}\sigma_{Y_{P^T}}}},\\
     d_{2}=\frac{\ln \frac{S_t}{\sqrt[n]{K}P(t,T)}-\int_{t}^{T}q(s)ds-\frac{1}{2}\sigma_{ Y_{P^T}}^2- \frac{1}{2}\sigma_{X_{P^T}}^2-\rho^{'}_t \sigma_{ X_{P^T}} \sigma_{ Y_{P^T}}} {\sqrt{\sigma_{X_{P^T}}^{2}+\sigma_{Y_{P^T}}^{2}+2\rho_t^{'}\sigma_{X_{P^T}}\sigma_{Y_{P^T}}}}\\
     \rho_t^{'}= \frac{ \int_{t}^{T} \rho_u \sigma_{r}(u)\sigma_{u}m(u,T)du}{ \sigma_{X_{P^T}}\sigma_{Y_{P^T}}}
     \end{array}
     \end{equation}

where $N(x)$ denotes standard normal distribution. \\

\begin{proof} According to arbitrage risk-neutral theory, the price of  European power call option with  payoff function (3) at time $t$ is
\begin{equation}
\begin{array}{ll}
V_{c}(t,r_{t},S_{t},P(t,T),T,K)&=\mathbb{E^Q}[ (S_{T}^{n}-K)^{+} e^{-\int_{t}^{T} r(s)ds}|\mathcal{F}_{t}]\\
                      &=P(t,T)\mathbb{E}^{P^T} [ (S_{T}^{n}-K)^{+}|\mathcal{F}_{t}]\\
                      &=P(t,T)\mathbb{E}^{P^T} [S_{T}^{n} I_{\{S_{T}^{n}>K\}} |\mathcal{F}_{t}]
                        -P(t,T) K \mathbb{E}^{P^T}[I_{\{S_{T}^{n}>K\}}|\mathcal{F}_{t}]\\
                      &=P(t,T)\mathbb{E}^{P^T} [S_{T}^{n} I_{\{S_{T}>\sqrt[n]{K}\}} |\mathcal{F}_{t}]
                        -P(t,T) K {P^T}[S_{T}>\sqrt[n]{K}|\mathcal{F}_{t}]\\
                      &=P(t,T)\mathbb{E}^{P^T} [(\frac{S_{T}}{P(T,T)})^{n} I_{\{\frac{S_{T}}{P(T,T)}>\sqrt[n]{K}\}} |\mathcal{F}_{t}]
                        -P(t,T) K {P^T}[\frac{S_{T}}{P(T,T)}>\sqrt[n]{K}|\mathcal{F}_{t}]\\
  &\stackrel{\bigtriangleup}{=} A-B
\end{array}
\nonumber
\end{equation}

Since,
\begin{equation}
\rho_t^{'}=\frac{Cov(\frac{X_{P^T}}{\sigma_{X_{P^T}}},\frac{Y_{P^T}}{\sigma_{Y_{P^T}}})}{ \sqrt{D(\frac{X_{P^T}}{\sigma_{X_{P^T}}})}\sqrt{D(\frac{Y_{P^T}}{\sigma_{Y_{P^T}}}})}
         =\frac{ {Cov(X_{P^T},Y_{P^T})}}{ \sigma_{X_{P^T}}\sigma_{Y_{P^T}}}= \frac{ \int_{t}^{T}   \rho_u \sigma_{r}(u)\sigma_{S}(u)m(u,T)du}{ \sigma_{X_{P^T}}\sigma_{Y_{P^T}}}
\nonumber
\end{equation}
combining formulae (47)(48) and Lemma 4, we obtain

\begin{equation}
\begin{array}{ll}
A&=P(t,T)\mathbb{E}^{P^T} [(\frac{S_t}{P(t,T)})^n e^{n [\int_{t}^{T}(-q(s)-\frac{1}{2}\sigma_S^2(s)-\frac{1}{2}\sigma_r^2(s)m^2(s,T)-\rho_s \sigma_S(s)\sigma_r(s)m(s,T))ds+X_{P^T}+ Y_{P^T}]} \\
&\quad I_{\{\frac{S_t}{P(t,T)} e^{\int_{t}^{T}(-q(s)-\frac{1}{2}\sigma_S^2(s)-\frac{1}{2}\sigma_r^2(s)m^2(s,T)-\rho_s \sigma_S(s)\sigma_r(s)m(s,T))ds+X_{P^T}+ Y_{P^T}}>\sqrt[n]{K}\}} |\mathcal{F}_{t}]\\
&=P(t,T)\mathbb{E}^{P^T} [(\frac{S_t}{P(t,T)})^n e^{n [-\int_{t}^{T}q(s)ds -\frac{1}{2}\sigma_{ Y_{P^T}}^2-\frac{1}{2}\sigma_{X_{P^T}}^2-\rho^{'}_t \sigma_{ X_{P^T}} \sigma_{ Y_{P^T}} + X_{P^T}+ Y_{P^T}]} \\
&\quad I_{\{\frac{S_t}{P(t,T)} e^{-\int_{t}^{T}q(s) ds -\frac{1}{2}\sigma_{ Y_{P^T}}^2-\frac{1}{2}\sigma_{X_{P^T}}^2-\rho^{'}_t \sigma_{ X_{P^T}} \sigma_{ Y_{P^T}} + X_{P^T}+ Y_{P^T}}>\sqrt[n]{K} \}} |\mathcal{F}_{t}]\\
&=P(t,T)(\frac{S_t}{P(t,T)})^n  e^{n [-\int_{t}^{T}q(s) ds -\frac{1}{2}\sigma_{ Y_{P^T}}^2-\frac{1}{2}\sigma_{X_{P^T}}^2-\rho^{'}_t \sigma_{ X_{P^T}} \sigma_{ Y_{P^T}}]}
\mathbb{E}^{P^T} [ e^{n X_{P^T}+ nY_{P^T}} \\
&\quad I_{\{ X_{P^T}+ Y_{P^T}>\ln \frac{\sqrt[n]{K}P(t,T)}{S_t}+\int_{t}^{T}q(s)ds+\frac{1}{2}\sigma_{ Y_{P^T}}^2+ \frac{1}{2}\sigma_{X_{P^T}}^2+\rho^{'}_t \sigma_{ X_{P^T}} \sigma_{ Y_{P^T}}\}}|\mathcal{F}_{t}]\\
&=\frac{S^n_t}{P^{n-1}(t,T)} e^{n [-\int_{t}^{T}q(s) ds -\frac{1}{2}\sigma_{ Y_{P^T}}^2-\frac{1}{2}\sigma_{X_{P^T}}^2-\rho^{'}_t \sigma_{ X_{P^T}} \sigma_{ Y_{P^T}}]} \mathbb{E}^{P^T} [ e^{n X_{P^T}+ nY_{P^T}} \\
&\quad I_{\{ X_{P^T}+ Y_{P^T}>\ln \frac{\sqrt[n]{K} P(t,T)}{S_t}+\int_{t}^{T}q(s)ds+\frac{1}{2}\sigma_{ Y_{P^T}}^2+ \frac{1}{2}\sigma_{X_{P^T}}^2+\rho^{'}_t \sigma_{ X_{P^T}} \sigma_{Y_{P^T}}\}}|\mathcal{F}_{t}]\\
&= \frac{S^n_t}{P^{n-1}(t,T)} e^{n [-\int_{t}^{T}q(s) ds -\frac{1}{2}\sigma_{ Y_{P^T}}^2-\frac{1}{2}\sigma_{X_{P^T}}^2-\rho^{'}_t \sigma_{ X_{P^T}} \sigma_{ Y_{P^T}}]} e^{\frac{1}{2}[n^2 \sigma_{X_{P^T}}^{2}+n^2 \sigma_{Y_{P^T}}^{2}+2\rho_t^{'}n^2 \sigma_{X_{P^T}}\sigma_{Y_{P^T}}]}\\
& \quad  N\Big( \frac{\ln\frac{S_{t}}{\sqrt[n]{K} P(t,T)}-\int_{t}^{T} q(s)ds +(n-\frac{1}{2})\sigma_{X_{P^T}}^{2}+(n-\frac{1}{2})\sigma_{Y_{P^T}}^{2}+(2n-1)\rho_t^{'}\sigma_{X_{P^T}}\sigma_{Y_{P^T}}} { \sqrt{\sigma_{X_{P^T}}^{2}+\sigma_{Y_{P^T}}^{2}+2\rho_t^{'}\sigma_{X_{P^T}}\sigma_{Y_{P^T}}}}\Big)\\
&= \frac{S^n_t}{P^{n-1}(t,T)} e^{\frac{n(n-1)}{2}\sigma_{X_{P^T}}^{2}+\frac{n(n-1)}{2} \sigma_{Y_{P^T}}^{2}+\rho_t^{'}n(n-1)\sigma_{X_{P^T}}\sigma_{Y_{P^T}}-n\int_{t}^{T} q(s)ds  }\\
& \quad  N\Big( \frac{\ln\frac{S_{t}}{\sqrt[n]{K}P(t,T)}-\int_{t}^{T} q(s)ds +(n-\frac{1}{2})\sigma_{X_{P^T}}^{2}+(n-\frac{1}{2})\sigma_{Y_{P^T}}^{2}+(2n-1)\rho_t^{'}\sigma_{X_{P^T}}\sigma_{Y_{P^T}}} { \sqrt{\sigma_{X_{P^T}}^{2}+\sigma_{Y_{P^T}}^{2}+2\rho_t^{'}\sigma_{X_{P^T}}\sigma_{Y_{P^T}}}}\Big)\\
&= \frac{S^n_t}{P^{n-1}(t,T)} e^{\frac{n(n-1)}{2}\sigma_{X_{P^T}}^{2}+\frac{n(n-1)}{2} \sigma_{Y_{P^T}}^{2}+\rho_t^{'}n(n-1)\sigma_{X_{P^T}}\sigma_{Y_{P^T}}-n\int_{t}^{T} q(s)ds  } N(d_1)\\
\end{array}
\nonumber
\end{equation}

\begin{equation}
\begin{array}{ll}
B&=P(t,T) K {P^T}[\frac{S_t}{P(t,T)} e^{\int_{t}^{T}(-q(s)-\frac{1}{2}\sigma_S^2(s)-\frac{1}{2}\sigma_r^2(s)m^2(s,T)-\rho_s \sigma_S(s)\sigma_r(s)m(s,T))ds+X_{P^T}+ Y_{P^T}}>\sqrt[n]{K}|\mathcal{F}_{t}]\\
&=P(t,T) K {P^T}[ X_{P^T}+ Y_{P^T}>\ln \frac{\sqrt[n]{K}P(t,T)}{S_t}+\int_{t}^{T}q(s)ds+\frac{1}{2}\sigma_{ Y_{P^T}}^2+ \frac{1}{2}\sigma_{X_{P^T}}^2+\rho^{'}_t \sigma_{ X_{P^T}} \sigma_{Y_{P^T}} |\mathcal{F}_{t}]\\
&=P(t,T) K {P^T}[- X_{P^T}- Y_{P^T}<-(\ln \frac{\sqrt[n]{K}P(t,T)}{S_t}+\int_{t}^{T}q(s)ds+\frac{1}{2}\sigma_{ Y_{P^T}}^2+ \frac{1}{2}\sigma_{X_{P^T}}^2+\rho^{'}_t \sigma_{ X_{P^T}} \sigma_{ Y_{P^T}}) |\mathcal{F}_{t}]\\
&=P(t,T) K {P^T}[\frac{- X_{P^T}- Y_{P^T}} {\sqrt{\sigma_{X_{P^T}}^{2}+\sigma_{Y_{P^T}}^{2}+2\rho_t^{'}\sigma_{X_{P^T}}\sigma_{Y_{P^T}}}}<\frac{-(\ln \frac{\sqrt[n]{K}P(t,T)}{S_t}+\int_{t}^{T}q(s)ds+\frac{1}{2}\sigma_{ Y_{P^T}}^2+ \frac{1}{2}\sigma_{X_{P^T}}^2+\rho^{'}_t \sigma_{ X_{P^T}} \sigma_{ Y_{P^T}})}{\sqrt{\sigma_{X_{P^T}}^{2}+\sigma_{Y_{P^T}}^{2}+2\rho_t^{'}\sigma_{X_{P^T}}\sigma_{Y_{P^T}}}} |\mathcal{F}_{t}]\\
&=P(t,T) K N\Big(\frac{\ln \frac{S_t}{\sqrt[n]{K} P(t,T)}-\int_{t}^{T}q(s)ds-\frac{1}{2}\sigma_{ Y_{P^T}}^2- \frac{1}{2}\sigma_{X_{P^T}}^2-\rho^{'}_t \sigma_{ X_{P^T}} \sigma_{ Y_{P^T}}} {\sqrt{\sigma_{X_{P^T}}^{2}+\sigma_{Y_{P^T}}^{2}+2\rho_t^{'}\sigma_{X_{P^T}}\sigma_{Y_{P^T}}}}  \Big)\\
&=P(t,T) K N(d_2)\\
\end{array}
\nonumber
\end{equation}

Hence, formulae (53) is proved. And, Formula (54) can be derived similarly. According to the property of normal distribution $N(-x)=1-N(x)$, formula (55) can be easily verified. Therefore, it completes the proof. \\
\end{proof}

 \section{Vasicek interest rate process and exponential Ornstein-Uhlenbeck process are in real--world probability space with numeraire change}

 {\bf Market Assumption 2} can be taken the following concise form:

\begin{equation}
\left\{\begin{array}{l}
      \displaystyle dS(t)=(r(t)-q(t)) S(t) dt +\sigma_{S}(t) S(t) dB^Q(t)\\
      \displaystyle d r_{t}= \displaystyle (\beta_{t}-\alpha_{t}r_{t})dt+ \rho_t \sigma_{r}(t)  d B^Q(t) + \sqrt{1-\rho_t^2} \sigma_{r}(t)  dB^{\perp,Q}(t)   \\
\end{array}\right.
\end{equation}
where $dB^Q(t) d B^{\perp,Q}(t)=0$. \\

Calculating $\int_{t}^{T}  r_{s} ds$ and $P(t,T)$, we obtain

 {\bf Lemma 9.} For Vasciek interest rate $r(t)$ in formula (57), $0\leq t\leq T$ ,
 \begin{equation}
 \begin{array}{ll}
\displaystyle \int_{t}^{T} r_{s} ds&=\displaystyle  r_{t}m(t,T) + \int_{t}^{T} \beta_{s} m(s,T) ds\\
                                    &+\int_{t}^{T} \rho_s \sigma_{r} (s) m(s,T) dB^Q(s) + \int_{t}^{T} \sqrt{1-\rho_s^2}  \sigma_{r} (s) m(s,T) dB^{\perp,Q}(s)  \\
\end{array}
\end{equation}

{\bf Lemma 10.}
\begin{equation}
 \begin{array}{ll}
\displaystyle P(t,T)&=\frac{1}{D(t)} \mathbb{E^Q} [D(T)|\mathcal{F}_{t}]=\mathbb{E^Q} [e^{-\int_{t}^{T} r_s ds}|\mathcal{F}_{t}]\\
                    &=\exp\{-r_t m(t,T)-\int_{t}^{T} \beta_s m(s,T)ds +\frac{1}{2} \int_{t}^{T} \sigma_r^2(s) m^2(s,T) ds  \} \\
\end{array}
\end{equation}

Applying Ito formula to $P(t,T)$, we obtain

\begin{equation}
\begin{array}{ll}
\displaystyle \frac{d P(t,T)}{ P(t,T)}=r_t dt-[\rho_t \sigma_{r}(t)  d B^Q(t) + \sqrt{1-\rho_t^2} \sigma_{r}(t)  dB^{\perp,Q}(t)]  \\
\end{array}
\end{equation}

{\bf Lemma 11.} For $P(t,T)$ with $r(t)$ in formula (57), $D(t)P(t,T)$ is a $Q$-martingale.

\begin{proof} Applying Ito formula,
\begin{equation}
\begin{array}{l}
\displaystyle d(D(t)P(t,T))=-D(t)P(t,T)[\rho_t \sigma_{r}(t)  d B^Q(t) + \sqrt{1-\rho_t^2} \sigma_{r}(t)  dB^{\perp,Q}(t)] .\\
\end{array}
\end{equation}
Hence, it is a $Q$-martingale.
\end{proof}

Thus, we obtain Radon-Nikodym derivatives of probability measure $ P^T $ with respect to $Q$,  the price of  payoff $V(T)$ at time t is
\begin{equation}
\begin{array}{ll}
\displaystyle V(t,r_{t},S_{t},P(t,T),T,K)=\displaystyle \frac{1}{D(t)} \mathbb{E^Q} [D(T)V(T)|\mathcal{F}_{t}]=\displaystyle P(t,T) \mathbb{E}^{P^T} [V(T)|\mathcal{F}_{t}] \\
\end{array}
\end{equation}

Let
\begin{equation}
\begin{array}{l}
\displaystyle B^{P^T}(t)=B^Q(t)+\int_{0}^{t} \rho_s \sigma_r(s)m(s,T) ds, \quad B^{\perp,P^T}(t)=B^{\perp,Q}(t)+\int_{0}^{t} \sqrt{1-\rho^2_s}\sigma_r(s)m(s,T) ds .\\
\end{array}
\end{equation}

According to Girsanov Theorem, $B^{P^T}(t)$ and $B^{\perp,P^T}(t)$ are independent Brownian motions under $P^T$.
Denote
\begin{equation}
\begin{array}{ll}
\tilde{X}_{P^T}=\int_{t}^{T} \rho_u \sigma_{r}(u) m(u,T)dB^{P^T}(u)+\int_{t}^{T}  \sqrt{1-\rho_u^2} \sigma_{r}(u)m(u,T)dB^{\perp,{P^T}}(u)\\
\tilde{Y}_{P^T}=\int_{t}^{T} \sigma_{S}(u)dB^{P^T}(u),\\
\end{array}
\end{equation}

We obtain
\begin{equation}
\begin{array}{ll}
\sigma_{\tilde{X}_{P^T}}^{2}=\int_{t}^{T} \sigma_{r}^{2}(u)m^{2}(u,T)du,  \quad \sigma_{\tilde{Y}_{P^T}}^{2}=\int_{t}^{T} \sigma_{S}^{2}(u)du    \\
\end{array}
\end{equation}
And,
\begin{equation}
\begin{array}{ll}
\displaystyle \frac{S(T)}{P(T,T)}=\displaystyle \frac{S(t)}{P(t,T)} e^{\int_{t}^{T}(-q(s)-\frac{1}{2}\sigma_S^2(s)-\frac{1}{2}\sigma_r^2(s)m^2(s,T)-\rho_s \sigma_S(s)\sigma_r(s)m(s,T))ds +\tilde{X}_{P^T}+ \tilde{Y}_{P^T}}.\\
\end{array}
\end{equation}

We obtain European power option formulae  as follows.\\

{\bf Theorem 7 } Suppose $Z(t)$ and $B(t)$  are two standard one-dimension Brown motions on probability space with correlation coefficient $\rho_t$ $( -1\leq\rho_t \leq 1)$, interest rate and stock process satisfy stochastic differential equations (19)(21)(24)(57), with strike price $K$, and maturity time $T$, European power call option price $V_{c}(t,r_{t},S_{t},P(t,T),T,K)$ at time $t (0\leq t\leq T)$ with payoff function (2) is

\begin{equation}
\begin{array}{ll}
V_{c}(t,r_{t},S_{t},P(t,T),T,K)=\frac{S^n_t}{P^{n-1}(t,T)} e^{\frac{n(n-1)}{2}\sigma_{\tilde{X}_{P^T}}^{2}+\frac{n(n-1)}{2} \sigma_{\tilde{Y}_{P^T}}^{2}+\rho_t^{'}n(n-1)\sigma_{\tilde{X}_{P^T}}\sigma_{\tilde{Y}_{P^T}}-n\int_{t}^{T} q(s)ds  }N( d_1)\\
\hspace*{2.0cm}-P(t,T) K^{n}N(d_2) \\
\end{array}
\end{equation}

European power put option price $V_{p}(t,r_{t},S_{t},P(t,T),T,K)$ at time $t (0\leq t\leq T)$ with payoff function (4) is

\begin{equation}
     \begin{array}{ll}
      V_{p}(t,r_{t},S_{t},P(t,T),T,K)=P(t,T) K^{n}N(-d_2)\\
                         \hspace*{2.0cm} -\frac{S^n_t}{P^{n-1}(t,T)} e^{\frac{n(n-1)}{2}\sigma_{\tilde{X}_{P^T}}^{2}+\frac{n(n-1)}{2} \sigma_{\tilde{Y}_{P^T}}^{2}+\rho_t^{'}n(n-1)\sigma_{\tilde{X}_{P^T}}\sigma_{\tilde{Y}_{P^T}}-n\int_{t}^{T} q(s)ds  } N(-d_1)\\
     \end{array}
     \end{equation}

European power put--call parity is

 \begin{equation}
    \begin{array}{ll}
    &V_{c}(t,r_{t},S_{t},P(t,T),T,K)+P(t,T) K^{n}\\
    &=V_{p}(t,r_{t},S_{t},P(t,T),T,K)+\frac{S^n_t}{P^{n-1}(t,T)} e^{\frac{n(n-1)}{2}\sigma_{\tilde{X}_{P^T}}^{2}+\frac{n(n-1)}{2} \sigma_{\tilde{Y}_{P^T}}^{2}+\rho_t^{'}n(n-1)\sigma_{\tilde{X}_{P^T}}\sigma_{\tilde{Y}_{P^T}}-n\int_{t}^{T} q(s)ds  }
     \end{array}
    \end{equation}
where,
     \begin{equation}
     \begin{array}{ll}
     d_{1}=\frac{\ln\frac{S_{t}}{KP(t,T)}-\int_{t}^{T} q(s)ds +(n-\frac{1}{2})\sigma_{\tilde{X}_{P^T}}^{2}+(n-\frac{1}{2})\sigma_{\tilde{Y}_{P^T}}^{2}+(2n-1)\rho_t^{'}\sigma_{\tilde{X}_{P^T}}\sigma_{\tilde{Y}_{P^T}}} { \sqrt{\sigma_{\tilde{X}_{P^T}}^{2}+\sigma_{\tilde{Y}_{P^T}}^{2}+2\rho_t^{'}\sigma_{\tilde{X}_{P^T}}\sigma_{\tilde{Y}_{P^T}}}},\\
     d_{2}=\frac{\ln \frac{S_t}{KP(t,T)}-\int_{t}^{T}q(s)ds-\frac{1}{2}\sigma_{ \tilde{Y}_{P^T}}^2- \frac{1}{2}\sigma_{\tilde{X}_{P^T}}^2-\rho^{'}_t \sigma_{ \tilde{X}_{P^T}} \sigma_{ \tilde{Y}_{P^T}}} {\sqrt{\sigma_{\tilde{X}_{P^T}}^{2}+\sigma_{\tilde{Y}_{P^T}}^{2}+2\rho_t^{'}\sigma_{\tilde{X}_{P^T}}\sigma_{\tilde{Y}_{P^T}}}}\\
     \rho_t^{'}= \frac{ \int_{t}^{T} \rho_u \sigma_{r}(u)\sigma_{u}m(u,T)du}{ \sigma_{\tilde{X}_{P^T}}\sigma_{\tilde{Y}_{P^T}}}
     \end{array}
     \end{equation}

where $N(x)$ denotes standard normal distribution.\\

\begin{proof}  According to arbitrage risk-neutral theory, using T-forward measure with zero-coupon bond as a numeraire, European power call option at time $t$  with  payoff function $\xi=(S_{T}^{n}-K^{n})^{+}$ is
\begin{equation}
\begin{array}{ll}
V_{c}(t,r_{t},S_{t},P(t,T),T,K)&=\mathbb{E^Q}[ (S_{T}^{n}-K^{n})^{+} e^{-\int_{t}^{T} r(s)ds}|\mathcal{F}_{t}]\\
                      &=P(t,T)\mathbb{E}^{P^T} [ (S_{T}^{n}-K^{n})^{+}|\mathcal{F}_{t}]\\
                      &=P(t,T)\mathbb{E}^{P^T} [S_{T}^{n} I_{\{S_{T}^{n}>K^{n}\}} |\mathcal{F}_{t}]
                        -P(t,T) K^{n}\mathbb{E}^{P^T}[I_{\{S_{T}^{n}>K^{n}\}}|\mathcal{F}_{t}]\\
                      &=P(t,T)\mathbb{E}^{P^T} [S_{T}^{n} I_{\{S_{T}>K\}} |\mathcal{F}_{t}]
                        -P(t,T) K^{n} {P^T}[S_{T}>K|\mathcal{F}_{t}]\\
                      &=P(t,T)\mathbb{E}^{P^T} [(\frac{S_{T}}{P(T,T)})^{n} I_{\{\frac{S_{T}}{P(T,T)}>K\}} |\mathcal{F}_{t}]
                        -P(t,T) K^{n} {P^T}[\frac{S_{T}}{P(T,T)}>K|\mathcal{F}_{t}]\\
                       &\stackrel{\bigtriangleup}{=} A-B
\end{array}
\nonumber
\end{equation}

Since,
\begin{equation}
\rho_t^{'}=\frac{Cov(\frac{\tilde{X}_{P^T}}{\sigma_{\tilde{X}_{P^T}}},\frac{\tilde{Y}_{P^T}}{\sigma_{\tilde{Y}_{P^T}}})}{ \sqrt{D(\frac{\tilde{X}_{P^T}}{\sigma_{\tilde{X}_{P^T}}})}\sqrt{D(\frac{\tilde{Y}_{P^T}}{\sigma_{\tilde{Y}_{P^T}}}})}
         =\frac{ {Cov(\tilde{X}_{P^T},\tilde{Y}_{P^T})}}{ \sigma_{\tilde{X}_{P^T}}\sigma_{\tilde{Y}_{P^T}}}= \frac{ \int_{t}^{T}   \rho_u \sigma_{r}(u)\sigma_{S}(u)m(u,T)du}{ \sigma_{\tilde{X}_{P^T}}\sigma_{\tilde{Y}_{P^T}}}
         \nonumber
         \end{equation}
combining formulae (62)(64)(65)(66) and Lemma 3, we obtain

\begin{equation}
\begin{array}{ll}
A&=P(t,T)\mathbb{E}^{P^T} [(\frac{S_t}{P(t,T)})^n e^{n [\int_{t}^{T}(-q(s)-\frac{1}{2}\sigma_S^2(s)-\frac{1}{2}\sigma_r^2(s)m^2(s,T)-\rho_s \sigma_S(s)\sigma_r(s)m(s,T))ds+\tilde{X}_{P^T}+ \tilde{Y}_{P^T}]} \\
&\quad I_{\{\frac{S_t}{P(t,T)} e^{\int_{t}^{T}(-q(s)-\frac{1}{2}\sigma_S^2(s)-\frac{1}{2}\sigma_r^2(s)m^2(s,T)-\rho_s \sigma_S(s)\sigma_r(s)m(s,T))ds+\tilde{X}_{P^T}+ \tilde{Y}_{P^T}}>K\}} |\mathcal{F}_{t}]\\
&=P(t,T)\mathbb{E}^{P^T} [(\frac{S_t}{P(t,T)})^n e^{n [-\int_{t}^{T}q(s)ds -\frac{1}{2}\sigma_{ \tilde{Y}_{P^T}}^2-\frac{1}{2}\sigma_{\tilde{X}_{P^T}}^2-\rho^{'}_t \sigma_{ \tilde{X}_{P^T}} \sigma_{ \tilde{Y}_{P^T}} + \tilde{X}_{P^T}+ \tilde{Y}_{P^T}]} \\
&\quad I_{\{\frac{S_t}{P(t,T)} e^{-\int_{t}^{T}q(s) ds -\frac{1}{2}\sigma_{ \tilde{Y}_{P^T}}^2-\frac{1}{2}\sigma_{\tilde{X}_{P^T}}^2-\rho^{'}_t \sigma_{ \tilde{X}_{P^T}} \sigma_{ \tilde{Y}_{P^T}} + \tilde{X}_{P^T}+ \tilde{Y}_{P^T}}>K \}} |\mathcal{F}_{t}]\\
&=P(t,T)(\frac{S_t}{P(t,T)})^n  e^{n [-\int_{t}^{T}q(s) ds -\frac{1}{2}\sigma_{ \tilde{Y}_{P^T}}^2-\frac{1}{2}\sigma_{\tilde{X}_{P^T}}^2-\rho^{'}_t \sigma_{ \tilde{X}_{P^T}} \sigma_{ \tilde{Y}_{P^T}}]}
\mathbb{E}^{P^T} [ e^{n \tilde{X}_{P^T}+ n\tilde{Y}_{P^T}} \\
&\quad I_{\{ \tilde{X}_{P^T}+ \tilde{Y}_{P^T}>\ln \frac{KP(t,T)}{S_t}+\int_{t}^{T}q(s)ds+\frac{1}{2}\sigma_{ \tilde{Y}_{P^T}}^2+ \frac{1}{2}\sigma_{\tilde{X}_{P^T}}^2+\rho^{'}_t \sigma_{ \tilde{X}_{P^T}} \sigma_{ \tilde{Y}_{P^T}}\}}|\mathcal{F}_{t}]\\
&=\frac{S^n_t}{P^{n-1}(t,T)} e^{n [-\int_{t}^{T}q(s) ds -\frac{1}{2}\sigma_{ \tilde{Y}_{P^T}}^2-\frac{1}{2}\sigma_{\tilde{X}_{P^T}}^2-\rho^{'}_t \sigma_{ \tilde{X}_{P^T}} \sigma_{ \tilde{Y}_{P^T}}]} \mathbb{E}^{P^T} [ e^{n \tilde{X}_{P^T}+ n\tilde{Y}_{P^T}} \\
&\quad I_{\{ \tilde{X}_{P^T}+ \tilde{Y}_{P^T}>\ln \frac{KP(t,T)}{S_t}+\int_{t}^{T}q(s)ds+\frac{1}{2}\sigma_{ \tilde{Y}_{P^T}}^2+ \frac{1}{2}\sigma_{\tilde{X}_{P^T}}^2+\rho^{'}_t \sigma_{ \tilde{X}_{P^T}} \sigma_{\tilde{Y}_{P^T}}\}}|\mathcal{F}_{t}]\\
&= \frac{S^n_t}{P^{n-1}(t,T)} e^{n [-\int_{t}^{T}q(s) ds -\frac{1}{2}\sigma_{ \tilde{Y}_{P^T}}^2-\frac{1}{2}\sigma_{\tilde{X}_{P^T}}^2-\rho^{'}_t \sigma_{ \tilde{X}_{P^T}} \sigma_{ \tilde{Y}_{P^T}}]} e^{\frac{1}{2}[n^2 \sigma_{\tilde{X}_{P^T}}^{2}+n^2 \sigma_{\tilde{Y}_{P^T}}^{2}+2\rho_t^{'}n^2 \sigma_{\tilde{X}_{P^T}}\sigma_{\tilde{Y}_{P^T}}]}\\
& \quad  N\Big( \frac{\ln\frac{S_{t}}{KP(t,T)}-\int_{t}^{T} q(s)ds +(n-\frac{1}{2})\sigma_{\tilde{X}_{P^T}}^{2}+(n-\frac{1}{2})\sigma_{\tilde{Y}_{P^T}}^{2}+(2n-1)\rho_t^{'}\sigma_{\tilde{X}_{P^T}}\sigma_{\tilde{Y}_{P^T}}} { \sqrt{\sigma_{\tilde{X}_{P^T}}^{2}+\sigma_{\tilde{Y}_{P^T}}^{2}+2\rho_t^{'}\sigma_{\tilde{X}_{P^T}}\sigma_{\tilde{Y}_{P^T}}}}\Big)\\
&= \frac{S^n_t}{P^{n-1}(t,T)} e^{\frac{n(n-1)}{2}\sigma_{\tilde{X}_{P^T}}^{2}+\frac{n(n-1)}{2} \sigma_{\tilde{Y}_{P^T}}^{2}+\rho_t^{'}n(n-1)\sigma_{\tilde{X}_{P^T}}\sigma_{\tilde{Y}_{P^T}}-n\int_{t}^{T} q(s)ds  }\\
& \quad  N\Big( \frac{\ln\frac{S_{t}}{KP(t,T)}-\int_{t}^{T} q(s)ds +(n-\frac{1}{2})\sigma_{\tilde{X}_{P^T}}^{2}+(n-\frac{1}{2})\sigma_{\tilde{Y}_{P^T}}^{2}+(2n-1)\rho_t^{'}\sigma_{\tilde{X}_{P^T}}\sigma_{\tilde{Y}_{P^T}}} { \sqrt{\sigma_{\tilde{X}_{P^T}}^{2}+\sigma_{\tilde{Y}_{P^T}}^{2}+2\rho_t^{'}\sigma_{\tilde{X}_{P^T}}\sigma_{\tilde{Y}_{P^T}}}}\Big)\\
&= \frac{S^n_t}{P^{n-1}(t,T)} e^{\frac{n(n-1)}{2}\sigma_{\tilde{X}_{P^T}}^{2}+\frac{n(n-1)}{2} \sigma_{\tilde{Y}_{P^T}}^{2}+\rho_t^{'}n(n-1)\sigma_{\tilde{X}_{P^T}}\sigma_{\tilde{Y}_{P^T}}-n\int_{t}^{T} q(s)ds  } N(d_1)\\
\end{array}
\nonumber
\end{equation}

\begin{equation}
\begin{array}{ll}
B&=P(t,T) K^{n} {P^T}[\frac{S_t}{P(t,T)} e^{\int_{t}^{T}(-q(s)-\frac{1}{2}\sigma_S^2(s)-\frac{1}{2}\sigma_r^2(s)m^2(s,T)-\rho_s \sigma_S(s)\sigma_r(s)m(s,T))ds+\tilde{X}_{P^T}+ \tilde{Y}_{P^T}}>K |\mathcal{F}_{t}]\\
&=P(t,T) K^{n} {P^T}[ \tilde{X}_{P^T}+ \tilde{Y}_{P^T}>\ln \frac{KP(t,T)}{S_t}+\int_{t}^{T}q(s)ds+\frac{1}{2}\sigma_{ \tilde{Y}_{P^T}}^2+ \frac{1}{2}\sigma_{\tilde{X}_{P^T}}^2+\rho^{'}_t \sigma_{ \tilde{X}_{P^T}} \sigma_{ \tilde{Y}_{P^T}} |\mathcal{F}_{t}]\\
&=P(t,T) K^{n} {P^T}[- \tilde{X}_{P^T}- \tilde{Y}_{P^T}<-(\ln \frac{KP(t,T)}{S_t}+\int_{t}^{T}q(s)ds+\frac{1}{2}\sigma_{ \tilde{Y}_{P^T}}^2+ \frac{1}{2}\sigma_{\tilde{X}_{P^T}}^2+\rho^{'}_t \sigma_{ \tilde{X}_{P^T}} \sigma_{ \tilde{Y}_{P^T}}) |\mathcal{F}_{t}]\\
&=P(t,T) K^{n} {P^T}[\frac{- \tilde{X}_{P^T}- \tilde{Y}_{P^T}}{\sqrt{\sigma_{\tilde{X}_{P^T}}^{2}+\sigma_{\tilde{Y}_{P^T}}^{2}+2\rho_t^{'}\sigma_{\tilde{X}_{P^T}}\sigma_{\tilde{Y}_{P^T}}}}<\frac{-(\ln \frac{KP(t,T)}{S_t}+\int_{t}^{T}q(s)ds+\frac{1}{2}\sigma_{ \tilde{Y}_{P^T}}^2+ \frac{1}{2}\sigma_{\tilde{X}_{P^T}}^2+\rho^{'}_t \sigma_{ \tilde{X}_{P^T}} \sigma_{ \tilde{Y}_{P^T}})}{\sqrt{\sigma_{\tilde{X}_{P^T}}^{2}+\sigma_{\tilde{Y}_{P^T}}^{2}+2\rho_t^{'}\sigma_{\tilde{X}_{P^T}}\sigma_{\tilde{Y}_{P^T}}}} |\mathcal{F}_{t}]\\
&=P(t,T) K^{n}N\Big(\frac{\ln \frac{S_t}{KP(t,T)}-\int_{t}^{T}q(s)ds-\frac{1}{2}\sigma_{ \tilde{Y}_{P^T}}^2- \frac{1}{2}\sigma_{\tilde{X}_{P^T}}^2-\rho^{'}_t \sigma_{ \tilde{X}_{P^T}} \sigma_{ \tilde{Y}_{P^T}}} {\sqrt{\sigma_{\tilde{X}_{P^T}}^{2}+\sigma_{\tilde{Y}_{P^T}}^{2}+2\rho_t^{'}\sigma_{\tilde{X}_{P^T}}\sigma_{\tilde{Y}_{P^T}}}}  \Big)\\
&=P(t,T) K^{n}N(d_2)\\
\end{array}
\nonumber
\end{equation}
Hence formula (67) is proved. Similarly,  formula (68) can be derived.
According to the property of normal distribution $N(-x)=1-N(x)$, formula (69) can be easily verified. Therefore, it completes the proof. \\
\end{proof}

{\bf Theorem 8 } Suppose $Z(t)$ and $B(t)$  are two standard one-dimension Brownian motions with correlation coefficient $\rho_t$ $( -1\leq\rho_t \leq 1)$, interest rate and stock process satisfy stochastic differential equations (19)(21)(24)(57), with strike price $K$, and maturity time $T$, the European power call option price  $V_{c}(t,r_{t},S_{t},P(t,T),T,K)$ at time $t$ $(0\leq t\leq T)$ with payoff function (3) is

\begin{equation}
\begin{array}{ll}
V_{c}(t,r_{t},S_{t},P(t,T),T,K)=\frac{S^n_t}{P^{n-1}(t,T)} e^{\frac{n(n-1)}{2}\sigma_{\tilde{X}_{P^T}}^{2}+\frac{n(n-1)}{2} \sigma_{\tilde{Y}_{P^T}}^{2}+\rho_t^{'}n(n-1)\sigma_{\tilde{X}_{P^T}}\sigma_{\tilde{Y}_{P^T}}-n\int_{t}^{T} q(s)ds  }N( d_1)\\
\hspace*{2.0cm}-P(t,T) K N(d_2) \\
\end{array}
\end{equation}

European power put option $V_{p}(t,r_{t},S_{t},P(t,T),T,K)$ at time $t (0\leq t\leq T)$ with pay--off function (5) is
\begin{equation}
     \begin{array}{ll}
      V_{p}(t,r_{t},S_{t},P(t,T),T,K)=P(t,T) K N(-d_2)\\
                         \hspace*{2.0cm} -\frac{S^n_t}{P^{n-1}(t,T)} e^{\frac{n(n-1)}{2}\sigma_{\tilde{X}_{P^T}}^{2}+\frac{n(n-1)}{2} \sigma_{\tilde{Y}_{P^T}}^{2}+\rho_t^{'}n(n-1)\sigma_{\tilde{X}_{P^T}}\sigma_{\tilde{Y}_{P^T}}-n\int_{t}^{T} q(s)ds  } N(-d_1)\\
     \end{array}
     \end{equation}

European power option put--call parity is
\begin{equation}
    \begin{array}{ll}
    &V_{c}(t,r_{t},S_{t},P(t,T),T,K)+P(t,T) K\\
    &=V_{p}(t,r_{t},S_{t},P(t,T),T,K)+\frac{S^n_t}{P^{n-1}(t,T)} e^{\frac{n(n-1)}{2}\sigma_{\tilde{X}_{P^T}}^{2}+\frac{n(n-1)}{2} \sigma_{\tilde{Y}_{P^T}}^{2}+\rho_t^{'}n(n-1)\sigma_{\tilde{X}_{P^T}}\sigma_{\tilde{Y}_{P^T}}-n\int_{t}^{T} q(s)ds  }
     \end{array}
    \end{equation}
where,
     \begin{equation}
     \begin{array}{ll}
     d_{1}=\frac{\ln\frac{S_{t}}{\sqrt[n]{K} P(t,T)}-\int_{t}^{T} q(s)ds +(n-\frac{1}{2})\sigma_{\tilde{X}_{P^T}}^{2}+(n-\frac{1}{2})\sigma_{\tilde{Y}_{P^T}}^{2}+(2n-1)\rho_t^{'}\sigma_{\tilde{X}_{P^T}}\sigma_{\tilde{Y}_{P^T}}} { \sqrt{\sigma_{\tilde{X}_{P^T}}^{2}+\sigma_{\tilde{Y}_{P^T}}^{2}+2\rho_t^{'}\sigma_{\tilde{X}_{P^T}}\sigma_{\tilde{Y}_{P^T}}}},\\
     d_{2}=\frac{\ln \frac{S_t}{\sqrt[n]{K}P(t,T)}-\int_{t}^{T}q(s)ds-\frac{1}{2}\sigma_{ \tilde{Y}_{P^T}}^2- \frac{1}{2}\sigma_{\tilde{X}_{P^T}}^2-\rho^{'}_t \sigma_{ \tilde{X}_{P^T}} \sigma_{ \tilde{Y}_{P^T}}} {\sqrt{\sigma_{\tilde{X}_{P^T}}^{2}+\sigma_{\tilde{Y}_{P^T}}^{2}+2\rho_t^{'}\sigma_{\tilde{X}_{P^T}}\sigma_{\tilde{Y}_{P^T}}}}\\
     \rho_t^{'}= \frac{ \int_{t}^{T} \rho_u \sigma_{r}(u)\sigma_{u}m(u,T)du}{ \sigma_{\tilde{X}_{P^T}}\sigma_{\tilde{Y}_{P^T}}}
     \end{array}
     \end{equation}

where $N(x)$ denotes standard normal distribution. \\

\begin{proof} According to arbitrage risk-neutral theory, the price of  European power call option with  payoff function (3) at time $t$ is
\begin{equation}
\begin{array}{ll}
V_{c}(t,r_{t},S_{t},P(t,T),T,K)&=\mathbb{E^Q}[ (S_{T}^{n}-K)^{+} e^{-\int_{t}^{T} r(s)ds}|\mathcal{F}_{t}]\\
                      &=P(t,T)\mathbb{E}^{P^T} [ (S_{T}^{n}-K)^{+}|\mathcal{F}_{t}]\\
                      &=P(t,T)\mathbb{E}^{P^T} [S_{T}^{n} I_{\{S_{T}^{n}>K\}} |\mathcal{F}_{t}]
                        -P(t,T) K \mathbb{E}^{P^T}[I_{\{S_{T}^{n}>K\}}|\mathcal{F}_{t}]\\
                      &=P(t,T)\mathbb{E}^{P^T} [S_{T}^{n} I_{\{S_{T}>\sqrt[n]{K}\}} |\mathcal{F}_{t}]
                        -P(t,T) K {P^T}[S_{T}>\sqrt[n]{K}|\mathcal{F}_{t}]\\
                      &=P(t,T)\mathbb{E}^{P^T} [(\frac{S_{T}}{P(T,T)})^{n} I_{\{\frac{S_{T}}{P(T,T)}>\sqrt[n]{K}\}} |\mathcal{F}_{t}]
                        -P(t,T) K {P^T}[\frac{S_{T}}{P(T,T)}>\sqrt[n]{K}|\mathcal{F}_{t}]\\
  &\stackrel{\bigtriangleup}{=} A-B
\end{array}
\nonumber
\end{equation}

Since,
\begin{equation}
\rho_t^{'}=\frac{Cov(\frac{\tilde{X}_{P^T}}{\sigma_{\tilde{X}_{P^T}}},\frac{\tilde{Y}_{P^T}}{\sigma_{\tilde{Y}_{P^T}}})}{ \sqrt{D(\frac{\tilde{X}_{P^T}}{\sigma_{\tilde{X}_{P^T}}})}\sqrt{D(\frac{\tilde{Y}_{P^T}}{\sigma_{\tilde{Y}_{P^T}}}})}
         =\frac{ {Cov(\tilde{X}_{P^T},\tilde{Y}_{P^T})}}{ \sigma_{\tilde{X}_{P^T}}\sigma_{\tilde{Y}_{P^T}}}= \frac{ \int_{t}^{T}   \rho_u \sigma_{r}(u)\sigma_{S}(u)m(u,T)du}{ \sigma_{\tilde{X}_{P^T}}\sigma_{\tilde{Y}_{P^T}}},
\nonumber
\end{equation}
combining formulae (62)(64)(65)(66) and Lemma 3, we obtain

\begin{equation}
\begin{array}{ll}
A&=P(t,T)\mathbb{E}^{P^T} [(\frac{S_t}{P(t,T)})^n e^{n [\int_{t}^{T}(-q(s)-\frac{1}{2}\sigma_S^2(s)-\frac{1}{2}\sigma_r^2(s)m^2(s,T)-\rho_s \sigma_S(s)\sigma_r(s)m(s,T))ds+\tilde{X}_{P^T}+ \tilde{Y}_{P^T}]} \\
&\quad I_{\{\frac{S_t}{P(t,T)} e^{\int_{t}^{T}(-q(s)-\frac{1}{2}\sigma_S^2(s)-\frac{1}{2}\sigma_r^2(s)m^2(s,T)-\rho_s \sigma_S(s)\sigma_r(s)m(s,T))ds+\tilde{X}_{P^T}+ \tilde{Y}_{P^T}}>\sqrt[n]{K}\}} |\mathcal{F}_{t}]\\
&=P(t,T)\mathbb{E}^{P^T} [(\frac{S_t}{P(t,T)})^n e^{n [-\int_{t}^{T}q(s)ds -\frac{1}{2}\sigma_{ \tilde{Y}_{P^T}}^2-\frac{1}{2}\sigma_{\tilde{X}_{P^T}}^2-\rho^{'}_t \sigma_{ \tilde{X}_{P^T}} \sigma_{ \tilde{Y}_{P^T}} + \tilde{X}_{P^T}+ \tilde{Y}_{P^T}]} \\
&\quad I_{\{\frac{S_t}{P(t,T)} e^{-\int_{t}^{T}q(s) ds -\frac{1}{2}\sigma_{ \tilde{Y}_{P^T}}^2-\frac{1}{2}\sigma_{\tilde{X}_{P^T}}^2-\rho^{'}_t \sigma_{ \tilde{X}_{P^T}} \sigma_{ \tilde{Y}_{P^T}} + \tilde{X}_{P^T}+ \tilde{Y}_{P^T}}>\sqrt[n]{K} \}} |\mathcal{F}_{t}]\\
&=P(t,T)(\frac{S_t}{P(t,T)})^n  e^{n [-\int_{t}^{T}q(s) ds -\frac{1}{2}\sigma_{ \tilde{Y}_{P^T}}^2-\frac{1}{2}\sigma_{\tilde{X}_{P^T}}^2-\rho^{'}_t \sigma_{ \tilde{X}_{P^T}} \sigma_{ \tilde{Y}_{P^T}}]}
\mathbb{E}^{P^T} [ e^{n \tilde{X}_{P^T}+ n\tilde{Y}_{P^T}} \\
&\quad I_{\{ \tilde{X}_{P^T}+ \tilde{Y}_{P^T}>\ln \frac{\sqrt[n]{K}P(t,T)}{S_t}+\int_{t}^{T}q(s)ds+\frac{1}{2}\sigma_{ \tilde{Y}_{P^T}}^2+ \frac{1}{2}\sigma_{\tilde{X}_{P^T}}^2+\rho^{'}_t \sigma_{ \tilde{X}_{P^T}} \sigma_{ \tilde{Y}_{P^T}}\}}|\mathcal{F}_{t}]\\
&=\frac{S^n_t}{P^{n-1}(t,T)} e^{n [-\int_{t}^{T}q(s) ds -\frac{1}{2}\sigma_{ \tilde{Y}_{P^T}}^2-\frac{1}{2}\sigma_{\tilde{X}_{P^T}}^2-\rho^{'}_t \sigma_{ \tilde{X}_{P^T}} \sigma_{ \tilde{Y}_{P^T}}]} \mathbb{E}^{P^T} [ e^{n \tilde{X}_{P^T}+ n\tilde{Y}_{P^T}} \\
&\quad I_{\{ \tilde{X}_{P^T}+ \tilde{Y}_{P^T}>\ln \frac{\sqrt[n]{K} P(t,T)}{S_t}+\int_{t}^{T}q(s)ds+\frac{1}{2}\sigma_{ \tilde{Y}_{P^T}}^2+ \frac{1}{2}\sigma_{\tilde{X}_{P^T}}^2+\rho^{'}_t \sigma_{ \tilde{X}_{P^T}} \sigma_{\tilde{Y}_{P^T}}\}}|\mathcal{F}_{t}]\\
&= \frac{S^n_t}{P^{n-1}(t,T)} e^{n [-\int_{t}^{T}q(s) ds -\frac{1}{2}\sigma_{ \tilde{Y}_{P^T}}^2-\frac{1}{2}\sigma_{\tilde{X}_{P^T}}^2-\rho^{'}_t \sigma_{ \tilde{X}_{P^T}} \sigma_{ \tilde{Y}_{P^T}}]} e^{\frac{1}{2}[n^2 \sigma_{\tilde{X}_{P^T}}^{2}+n^2 \sigma_{\tilde{Y}_{P^T}}^{2}+2\rho_t^{'}n^2 \sigma_{\tilde{X}_{P^T}}\sigma_{\tilde{Y}_{P^T}}]}\\
& \quad  N\Big( \frac{\ln\frac{S_{t}}{\sqrt[n]{K} P(t,T)}-\int_{t}^{T} q(s)ds +(n-\frac{1}{2})\sigma_{\tilde{X}_{P^T}}^{2}+(n-\frac{1}{2})\sigma_{\tilde{Y}_{P^T}}^{2}+(2n-1)\rho_t^{'}\sigma_{\tilde{X}_{P^T}}\sigma_{\tilde{Y}_{P^T}}} { \sqrt{\sigma_{\tilde{X}_{P^T}}^{2}+\sigma_{\tilde{Y}_{P^T}}^{2}+2\rho_t^{'}\sigma_{\tilde{X}_{P^T}}\sigma_{\tilde{Y}_{P^T}}}}\Big)\\
&= \frac{S^n_t}{P^{n-1}(t,T)} e^{\frac{n(n-1)}{2}\sigma_{\tilde{X}_{P^T}}^{2}+\frac{n(n-1)}{2} \sigma_{\tilde{Y}_{P^T}}^{2}+\rho_t^{'}n(n-1)\sigma_{\tilde{X}_{P^T}}\sigma_{\tilde{Y}_{P^T}}-n\int_{t}^{T} q(s)ds  }\\
& \quad  N\Big( \frac{\ln\frac{S_{t}}{\sqrt[n]{K}P(t,T)}-\int_{t}^{T} q(s)ds +(n-\frac{1}{2})\sigma_{\tilde{X}_{P^T}}^{2}+(n-\frac{1}{2})\sigma_{\tilde{Y}_{P^T}}^{2}+(2n-1)\rho_t^{'}\sigma_{\tilde{X}_{P^T}}\sigma_{\tilde{Y}_{P^T}}} { \sqrt{\sigma_{\tilde{X}_{P^T}}^{2}+\sigma_{\tilde{Y}_{P^T}}^{2}+2\rho_t^{'}\sigma_{\tilde{X}_{P^T}}\sigma_{\tilde{Y}_{P^T}}}}\Big)\\
&= \frac{S^n_t}{P^{n-1}(t,T)} e^{\frac{n(n-1)}{2}\sigma_{\tilde{X}_{P^T}}^{2}+\frac{n(n-1)}{2} \sigma_{\tilde{Y}_{P^T}}^{2}+\rho_t^{'}n(n-1)\sigma_{\tilde{X}_{P^T}}\sigma_{\tilde{Y}_{P^T}}-n\int_{t}^{T} q(s)ds  } N(d_1)\\
\end{array}
\nonumber
\end{equation}

\begin{equation}
\begin{array}{ll}
B&=P(t,T) K {P^T}[\frac{S_t}{P(t,T)} e^{\int_{t}^{T}(-q(s)-\frac{1}{2}\sigma_S^2(s)-\frac{1}{2}\sigma_r^2(s)m^2(s,T)-\rho_s \sigma_S(s)\sigma_r(s)m(s,T))ds+\tilde{X}_{P^T}+ \tilde{Y}_{P^T}}>\sqrt[n]{K}|\mathcal{F}_{t}]\\
&=P(t,T) K {P^T}[ \tilde{X}_{P^T}+ \tilde{Y}_{P^T}>\ln \frac{\sqrt[n]{K}P(t,T)}{S_t}+\int_{t}^{T}q(s)ds+\frac{1}{2}\sigma_{ \tilde{Y}_{P^T}}^2+ \frac{1}{2}\sigma_{\tilde{X}_{P^T}}^2+\rho^{'}_t \sigma_{ \tilde{X}_{P^T}} \sigma_{\tilde{Y}_{P^T}} |\mathcal{F}_{t}]\\
&=P(t,T) K {P^T}[- \tilde{X}_{P^T}- \tilde{Y}_{P^T}<-(\ln \frac{\sqrt[n]{K}P(t,T)}{S_t}+\int_{t}^{T}q(s)ds+\frac{1}{2}\sigma_{ \tilde{Y}_{P^T}}^2+ \frac{1}{2}\sigma_{\tilde{X}_{P^T}}^2+\rho^{'}_t \sigma_{ \tilde{X}_{P^T}} \sigma_{ \tilde{Y}_{P^T}}) |\mathcal{F}_{t}]\\
&=P(t,T) K {P^T}[\frac{- \tilde{X}_{P^T}- \tilde{Y}_{P^T}} {\sqrt{\sigma_{\tilde{X}_{P^T}}^{2}+\sigma_{\tilde{Y}_{P^T}}^{2}+2\rho_t^{'}\sigma_{\tilde{X}_{P^T}}\sigma_{\tilde{Y}_{P^T}}}}<\frac{-(\ln \frac{\sqrt[n]{K}P(t,T)}{S_t}+\int_{t}^{T}q(s)ds+\frac{1}{2}\sigma_{ \tilde{Y}_{P^T}}^2+ \frac{1}{2}\sigma_{\tilde{X}_{P^T}}^2+\rho^{'}_t \sigma_{ \tilde{X}_{P^T}} \sigma_{ \tilde{Y}_{P^T}})}{\sqrt{\sigma_{\tilde{X}_{P^T}}^{2}+\sigma_{\tilde{Y}_{P^T}}^{2}+2\rho_t^{'}\sigma_{\tilde{X}_{P^T}}\sigma_{\tilde{Y}_{P^T}}}} |\mathcal{F}_{t}]\\
&=P(t,T) K N\Big(\frac{\ln \frac{S_t}{\sqrt[n]{K} P(t,T)}-\int_{t}^{T}q(s)ds-\frac{1}{2}\sigma_{ \tilde{Y}_{P^T}}^2- \frac{1}{2}\sigma_{\tilde{X}_{P^T}}^2-\rho^{'}_t \sigma_{ \tilde{X}_{P^T}} \sigma_{ \tilde{Y}_{P^T}}} {\sqrt{\sigma_{\tilde{X}_{P^T}}^{2}+\sigma_{\tilde{Y}_{P^T}}^{2}+2\rho_t^{'}\sigma_{\tilde{X}_{P^T}}\sigma_{\tilde{Y}_{P^T}}}}  \Big)\\
&=P(t,T) K N(d_2)\\
\end{array}
\nonumber
\end{equation}
Hence, formulae (71) is proved. And, Formula (72) can be derived similarly. According to the property of normal distribution $N(-x)=1-N(x)$, formula (73) can be easily verified. Therefore, it completes the proof. \\
\end{proof}

\section{Conclusion}
\label{}

We develop two cases of European power option pricing under two assumptions of Vasciek interest rate process and exponential Ornstein-Uhlenbeck process, where Vasciek interest rate process is under equivalent martingale measurement derived by stock exponential Ornstein-Uhlenbeck process, or Vasciek interest rate process and  exponential Ornstein-Uhlenbeck stock process are in the real--world probability space. Then, we extend above conclusions under numeraire change.Finally, we solve the European power option pricing with Vasciek interest rate process and exponential Ornstein-Uhlenbeck process in real--world probability space, which are the theoretical foundation to practice and simulation. In fact, we develop European power option pricing under both money market account and forward numeraire. The future work will focus on the realization of European power option pricing in practice of real-world and extend our theory to other types of option pricing.

\end{document}